\documentclass[12pt]{article}
\usepackage{amsmath}
\usepackage{amssymb}
\usepackage{graphicx}
\usepackage{subfigure}
\usepackage{url}
\usepackage[colorlinks=true, linkcolor=blue, citecolor=blue,
urlcolor=blue]{hyperref}
\begin{document}
\baselineskip 0.6cm

\def\simprop{\mathrel{\lower3.0pt\vbox{\lineskip=1.0pt\baselineskip=0pt
             \hbox{$\propto$}\hbox{$\sim$}}}}

\begin{titlepage}

\begin{flushright}
UCB-PTH-14/07
\end{flushright}

\vskip 1.9cm

\begin{center}
{\Large \bf Grand Unification, Axion, and Inflation in \\
Intermediate Scale Supersymmetry}

\vskip 0.7cm

{\large Lawrence J. Hall, Yasunori Nomura, and Satoshi Shirai}

\vskip 0.4cm
{\it Berkeley Center for Theoretical Physics, Department of Physics, \\
     and Theoretical Physics Group, Lawrence Berkeley National Laboratory, \\
     University of California, Berkeley, CA 94720, USA} \\

\vskip 0.8cm

\abstract{A class of supersymmetric grand unified theories is introduced 
 that has a single scale below the cutoff, that of the supersymmetry 
 breaking masses $\tilde{m}$.  For a wide range of the dimensionless 
 parameters, agreement with the observed mass of the Higgs boson determines 
 $\tilde{m} \sim 10^9\mbox{--}10^{13}~{\rm GeV}$, yielding Intermediate 
 Scale Supersymmetry.  We show that within this framework it is possible 
 for seesaw neutrino masses, axions, and inflation to be described by 
 the scale $\tilde{m}$, offering the possibility of a unified origin of 
 disparate phenomena.  Neutrino masses allowing for thermal leptogenesis 
 can be obtained, and the axion decay constant lies naturally in the 
 range $f_a \sim 10^9\mbox{--}10^{11}~{\rm GeV}$, consistent with a recent 
 observational suggestion of high scale inflation.  A minimal $SU(5)$ 
 model is presented that illustrates these features.  In this model, 
 the only states at the grand unified scale are those of the heavy gauge 
 supermultiplet.  The grand unified partners of the Higgs doublets have 
 a mass of order $\tilde{m}$, leading to the dominant proton decay mode 
 $p \rightarrow \bar{\nu} K^+$, which may be probed in upcoming experiments. 
 Dark matter may be winos, with mass environmentally selected to the 
 TeV scale, and/or axions.  Gauge coupling unification is found to be 
 successful, especially if the wino is at the TeV scale.}

\end{center}
\end{titlepage}

\section{Introduction}
\label{sec:intro}

The key discovery from the first run of the Large Hadron Collider (LHC) 
is a highly perturbative Higgs boson coupled with no sign of any new 
physics that would allow a natural electroweak scale.  Remarkably, the 
value of the Higgs mass implies that the Standard Model (SM) remains 
perturbative to very high energy scales.  Although this ``Lonely Higgs'' 
picture could easily be overturned by discoveries at the next run of 
the LHC, at present we are confronted with a very surprising situation. 
A variety of new physics possibilities was introduced in the 1970s and 
1980s yielding a standard paradigm of a natural weak scale that was 
almost universally accepted.  While the absence of new physics at LEP 
and elsewhere led to doubts about naturalness, the Lonely Higgs discovery 
at LHC warrants new thinking on the naturalness of the weak scale and 
the likely mass scale of new physics.

An intriguing feature of the Lonely Higgs discovery is that the Higgs 
quartic coupling, on evolution to high energies, passes through zero and 
then remains close to zero up to unified scales, providing evidence for 
a highly perturbative Higgs sector at high energies.  This closeness to 
zero of the quartic coupling cannot be explained by the SM, and hence is 
a guide in seeking new physics at very high scales.  The Higgs boson mass 
was predicted to be in the range $\approx (128-141)~{\rm GeV}$ from a 
supersymmetric boundary condition at unified energies~\cite{Hall:2009nd}. 
Furthermore, it was pointed out that in such theories $\tan\beta$ 
near unity can result naturally, leading to a Higgs mass prediction 
of $(128 \pm 3)~{\rm GeV}$, with the central value gradually 
decreasing as the scale of supersymmetry is lowered below the unified 
scale.  After the Higgs boson discovery, the connection between 
supersymmetry at a high scale and the Higgs mass was investigated 
further~\cite{Hebecker:2012qp,Ibanez:2012zg}.

\begin{figure}[t!]
\begin{center}
\subfigure[Higgs quartic coupling $\lambda$]
 {\includegraphics[clip, width = 0.48 \textwidth]{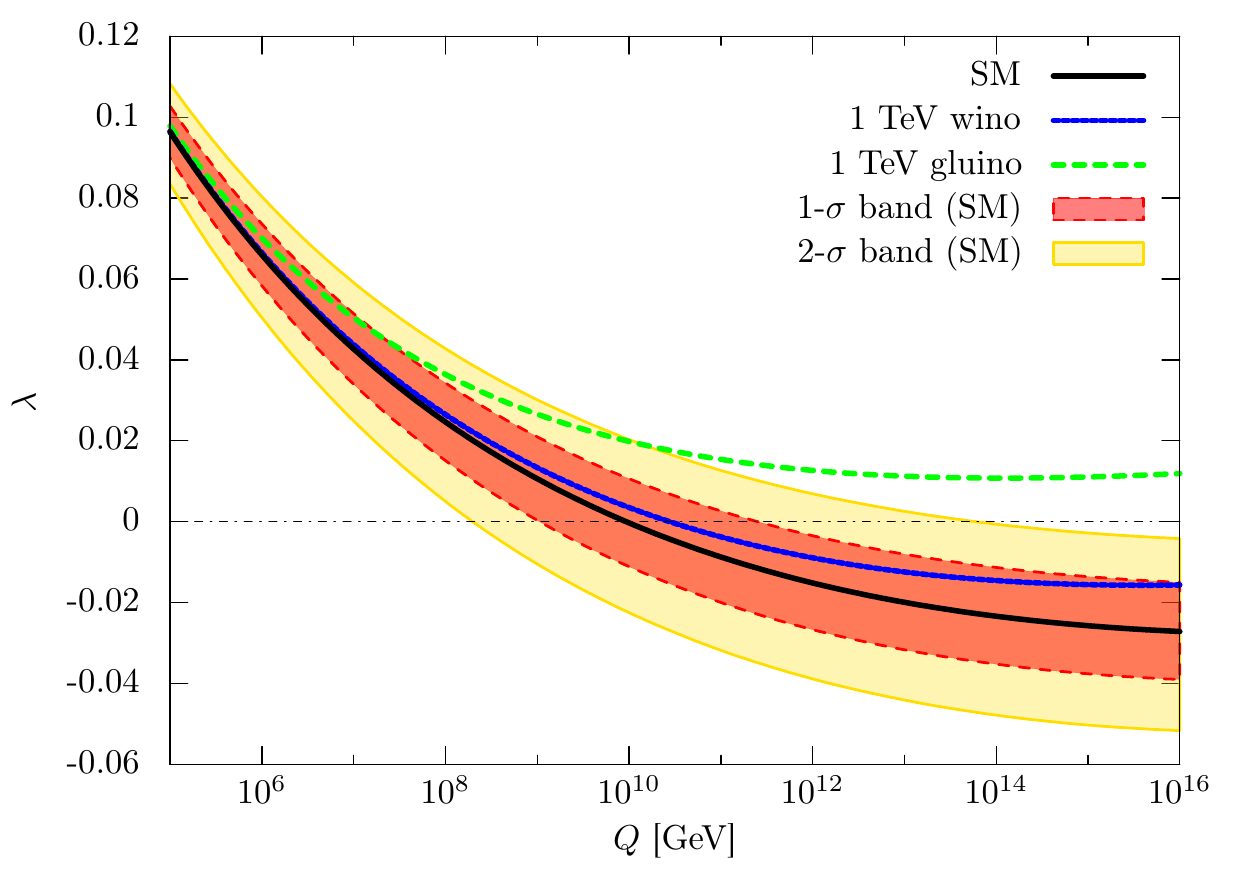}
 \label{fig:scale_lambda}}
\subfigure[1-$\sigma$ band of $\tan\beta$]
 {\includegraphics[clip, width = 0.48 \textwidth]{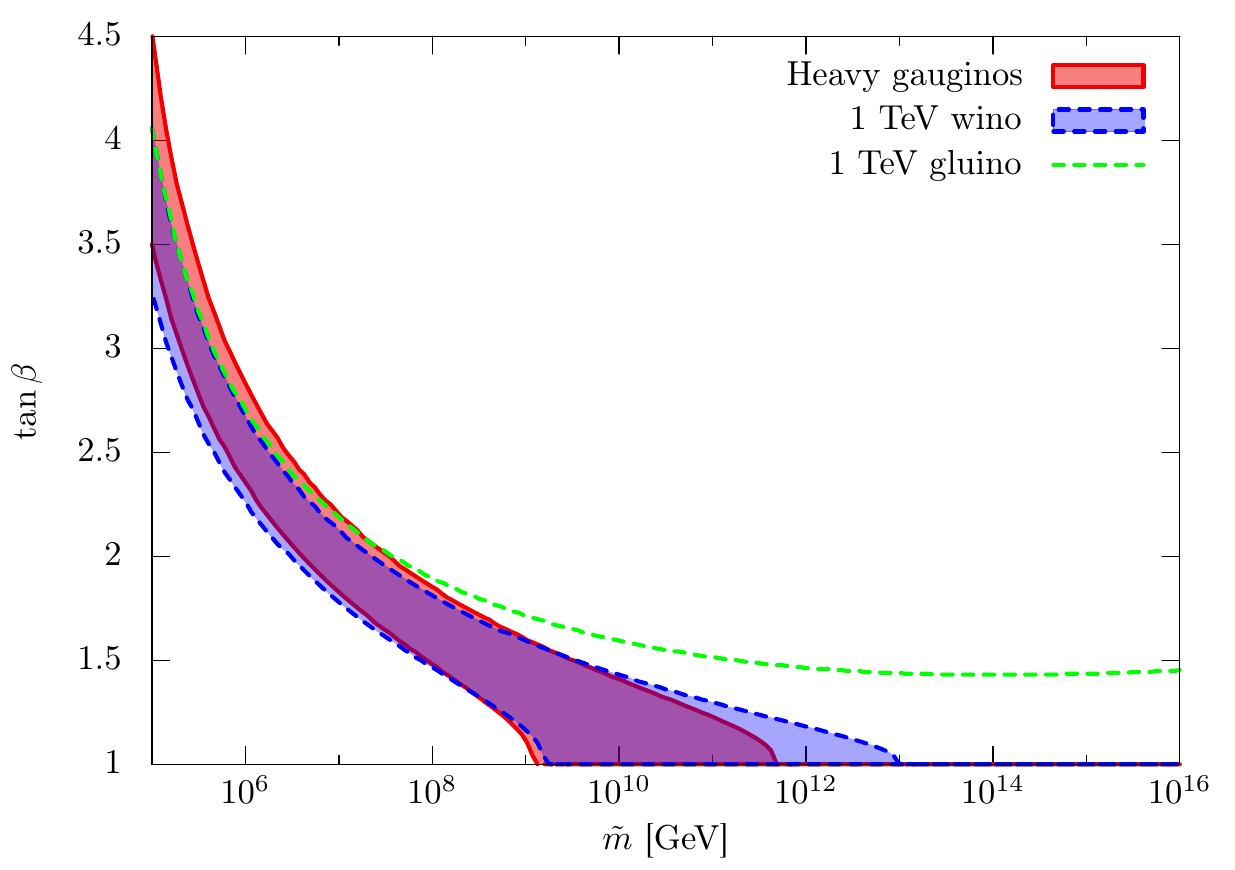}
 \label{fig:scale_tanb}}
\caption{(a) The renormalization group running of the Higgs quartic coupling 
 $\lambda$ for the SM (solid black line, with 1-$\sigma$ and 2-$\sigma$ 
 regions from uncertainties of the experimental input parameters indicated 
 by dark and light shades, respectively) and for the $1~{\rm TeV}$ wino 
 (solid blue) and gluino (dashed green) in addition to the SM particles. 
 (b) The value of $\tan\beta$ required to reproduce the observed Higgs 
 boson mass as a function of the superpartner mass scale $\tilde{m}$ in 
 the case that the theory below $\tilde{m}$ is the SM (red region bounded 
 by solid lines), the SM with $1~{\rm TeV}$ wino (blue region bounded 
 by dashed lines), and the SM with $1~{\rm TeV}$ gluino (dashed green 
 line).  The regions for the first two cases correspond to the 1-$\sigma$ 
 uncertainties for the input experimental parameters.}
\label{fig:scale}
\end{center}
\end{figure}

In a previous paper~\cite{Hall:2013eko}, two of us introduced {\it 
Intermediate Scale Supersymmetry} (ISS) to explain two key observations
\begin{itemize}
\item
The SM quartic coupling, when evolved to large scales, passes through 
zero at $\mu_c$.  This can be accounted for by taking the SM superpartner 
mass scale $\tilde{m} \sim \mu_c$.  From Fig.~\ref{fig:scale_lambda}, 
$\mu_c \sim 10^9~\mbox{--}10^{13}~{\rm GeV}$ at 1-$\sigma$ (allowing 
for the possibility of a TeV-scale wino for dark matter).
\item
States of a minimal supersymmetric grand unified theory at $\tilde{m}$ 
can account for precision gauge coupling unification.
\end{itemize}
In addition to these, in this paper we study ISS models that have a third 
key feature
\begin{itemize}
\item
Below the cutoff scale of the theory $\Lambda$, which is likely close to 
the Planck mass, the theory possesses only a single mass scale, $\tilde{m}$.
\end{itemize}
In this paper we study two different aspects of ISS.  In 
Section~\ref{sec:model}, we pursue a class of ISS models that 
lead more cleanly to the vanishing of the quartic coupling near 
$\tilde{m}$, have a new proton decay signal and are more elegant. 
In Section~\ref{sec:scales}, we argue that in ISS the mass scale 
$\tilde{m}$ may be identified with one or more key mass scales 
of new physics:\ the axion decay constant, the energy scale of 
inflation, and the seesaw scale for neutrino masses.

ISS provides a unifying theme to the diverse physics that we discuss, 
since it is all triggered by the same underlying mass scale.  The scale 
$\tilde{m}$ directly gives the superpartner masses and can also be the 
origin of the axion decay constant, inflation, and right-handed neutrino 
masses.  Within this framework, the scale of weak interactions and of 
the Grand Unified Theory (GUT) need some explanation.

In ISS the weak scale is highly fine-tuned, for example by twenty 
orders of magnitude for $\tilde{m} = 10^{12}~{\rm GeV}$, and can be 
understood in the multiverse, which provides a coherent framework for 
understanding both the fine-tuning of the weak scale and the cosmological 
constant~\cite{Weinberg:1987dv,Agrawal:1997gf}.  In the $SU(5)$ unified 
model introduced in this paper, the fields responsible for weak breaking, 
$H, \bar{H}$, and $SU(5)$ breaking, $\Sigma$, do not have supersymmetric 
mass terms and are massless in the supersymmetric flat-space limit. 
Once supersymmetry is broken and the cosmological constant is fine-tuned, 
supergravity interactions induce an effective superpotential
\begin{equation}
  W_{\rm eff} \sim \tilde{m} \times 
    \left( \Sigma^2, \; H \bar{H}, \; \frac{1}{\Lambda} \Sigma^3, \; 
    \frac{1}{\Lambda} H \Sigma \bar{H}, \; ...  \right).
\label{eq:W_effintro}
\end{equation}
This yields an $SU(5)$ breaking vacuum $\langle \Sigma \rangle \sim 
O(\Lambda)$, and we choose order unity coefficients so that this vacuum 
expectation value (VEV) is somewhat less than $\Lambda$.  The heavy 
$XY$ gauge supermultiplet lies just below $\Lambda$, while all other 
states in $H, \bar{H}, \Sigma$ have masses of order $\tilde{m}$.  These 
states make a significant contribution to gauge coupling unification, 
and the color triplet states in $H, \bar{H}$ yield an interesting 
proton decay signal.

The $H \Sigma \bar{H}$ coupling is of order $\tilde{m}/\Lambda$, and 
hence leads to a negligible contribution to the Higgs quartic, which 
is dominated by the electroweak gauge contribution:\ $\lambda(\tilde{m}) 
\simeq 0.03\, (\tan^2\!\beta -1)^2$ for $|\tan^2\!\beta - 1| \ll 1$, 
where $\lambda$ is normalized such that $V(h) \supset (\lambda/2) 
(h^\dagger h)^2$, and the angle $\beta$ defines the combination of 
Higgs doublets that is fine-tuned light to become the SM Higgs.  A 
value of $\tan^2\!\beta$ in the range of about $0.5$ to $2$ is sufficient 
to understand a small value of $\lambda(\tilde{m})$; however, in the 
limit that the Higgs mixing parameter (the Higgsino mass) $\mu$ becomes 
larger than $\tilde{m}$, $\tan^2\!\beta - 1 \sim O(\tilde{m}^2/\mu^2)$, 
so that $\lambda(\tilde{m})$ rapidly drops below $0.01$.%
\footnote{If $\mu$ is too large ($\mu/m_{\tilde{t}} \gtrsim 4$), 
 however, there can be sizable threshold corrections to $\lambda$ 
 which affect the relation between $\tilde{m}$ and $\tan\beta$ 
 in Fig.~\ref{fig:scale_tanb}.}

The organization of the rest of the paper is as follows.  In 
Section~\ref{sec:ISS} we closely examine the running of the Higgs 
quartic coupling in the SM, and with the addition of a TeV-scale wino, 
to determine the range of $\mu_c$.  In Section~\ref{sec:model} we 
introduce and study a specific simple $SU(5)$ GUT that is representative 
of a class of grand unified theories that have just a single mass 
scale, $\tilde{m}$.  We study the spectrum, dark matter, gauge coupling 
unification and proton decay in this model.  In Section~\ref{sec:scales} 
we argue that in ISS other fundamental physics may be linked to the 
scale $\tilde{m}$, in particular, neutrino masses, axions, and inflation.
Finally we summarize in Section~\ref{sec:summary}.

\section{Higgs Quartic Coupling and ISS}
\label{sec:ISS}

Before entering into the main part of the paper, in this section we discuss 
the scale of supersymmetry breaking suggested by the current experimental 
data within the ISS framework.  In Fig.~\ref{fig:scale_lambda}, we show 
the running of the $\overline{\rm MS}$ Higgs quartic coupling $\lambda$ 
in the SM as a function of the renormalization scale $Q$ (solid, black 
line).  Here, the dark and light shaded regions correspond to the 1-$\sigma$ 
and 2-$\sigma$ ranges for the experimental input parameters, respectively, 
for which we have used $m_t = 173.34(76)~{\rm GeV}$~\cite{ATLAS:2014wva}, 
$m_{h^0} = 125.40(45)~{\rm GeV}$~\cite{ATLAS:2013mma,Chatrchyan:2013lba}, 
$m_W = 80.367(7)~{\rm GeV}$, $m_Z = 91.1875(21)~{\rm GeV}$~\cite{Baak:2013ppa}, 
and $\alpha_3(m_Z) = 0.1184(7)$~\cite{Bethke:2012jm}.  The figure also 
shows the running of $\lambda$ in the cases that the wino and gluino 
exist at $1~{\rm TeV}$ in addition to the SM particles (solid blue and 
dashed green lines, respectively).  In drawing these lines/regions, we 
have used, following Ref.~\cite{Giudice:2011cg}, 2-loop (1-loop) threshold 
corrections and 3-loop (2-loop) renormalization group equations for 
the SM particles (the wino and gluino).  As can be seen from the 
figure, $\lambda$ crosses zero at an intermediate scale $\mu_c \sim 
10^9\mbox{--}10^{13}~{\rm GeV}$ for the SM, although uncertainties from 
experimental input parameters are still very large.  The situation is 
similar if there is a wino at $1~{\rm TeV}$ (which is not entirely trivial 
as the crossing scale is highly sensitive to physics at lower energies 
as can be seen in the case in which the gluino exists at $1~{\rm TeV}$).

In the rest of the paper, we assume that the Higgs quartic coupling indeed 
crosses zero at $\mu_c$ if we evolve it to higher energy scales using the SM 
renormalization group equations (or those with a TeV-scale wino), although 
the possibility of the crossing scale being around the unification/Planck 
scale $\sim 10^{16}~\mbox{--}~10^{18}~{\rm GeV}$ is not yet excluded if 
we allow 2- to 3-$\sigma$ ranges for the current experimental errors. 
As discussed above and in Ref.~\cite{Hall:2013eko}, we identify this scale 
to be the scale for the superpartner masses $\tilde{m}$, at which the 
supersymmetric standard model (together with a part of the GUT particles) 
is reduced to the SM (possibly together with a wino or Higgsino) at lower 
energies.  Since $\lambda \ll 1$ at this scale, this implies $\tan\beta 
\sim 1$.  In Fig.~\ref{fig:scale_tanb}, we show the value of $\tan\beta$ 
needed to reproduce the observed Higgs boson mass $m_{h^0} \simeq 
125~{\rm GeV}$ as a function of the superpartner mass scale.  Here, 
we have assumed that all the scalar superpartners have common mass 
$\tilde{m}$; the gaugino and Higgsino masses are also taken to be 
$\tilde{m}$ and the scalar trilinear $A$-terms are set to be zero. 
We find that for the cases of the SM and the SM with a TeV-scale wino, 
the superpartner masses must be at an intermediate scale:
\begin{equation}
  \tilde{m} \sim 10^9\mbox{--}10^{13}~{\rm GeV},
\label{eq:m-tilde}
\end{equation}
for $\tan\beta \sim 1$ at the 1-$\sigma$ level.%
\footnote{At the 2-$\sigma$ level, the region with $\tan\beta = 1$ reaches 
 the conventional GUT scale, $\sim 10^{16}~{\rm GeV}$, for the case with 
 a TeV-scale wino, so that this becomes the ``$\mbox{SM} + \tilde{w}$'' 
 case of Ref.~\cite{Elor:2009jp}.  It is interesting to see how future 
 refinements of experimental determinations of, e.g.\ $m_t$ and $\alpha_3$, 
 imply about the scale in which $\lambda$ crosses zero.}
As can be seen from the figure, and emphasized in Ref.~\cite{Hall:2013eko}, 
this conclusion does not require $\tan\beta$ to be extremely close to $1$. 
Indeed, for $0.5 \lesssim \tan^2\!\beta \lesssim 2$, the range of the 
superpartner masses suggested by the central values for the experimental 
data is still close to Eq.~(\ref{eq:m-tilde}), and there is a wide range 
of parameter space which leads to these values of $\tan\beta$.  (In fact, 
even values of $\tan\beta$ very close to $1$ can naturally be obtained 
if $\mu$ is somewhat larger than $\tilde{m}$.)  Below, we will use values 
of $\tilde{m}$ in Eq.~(\ref{eq:m-tilde}) as our guide in discussing 
ISS theories.

In this paper, we mostly assume that supersymmetry breaking is mediated 
to the GUT sector (the sector charged under the SM gauge group) at a 
high scale $M_*$ close to the UV cutoff scale $\Lambda$ of the unified 
theory, which we expect to be within an order of magnitude of the reduced 
Planck scale $M_{\rm Pl} \simeq 2.4 \times 10^{18}~{\rm GeV}$:\ $M_* 
\sim \Lambda \sim M_{\rm Pl}$.  We then find that the gravitino mass 
$m_{3/2} = F/\sqrt{3} M_{\rm Pl}$ is roughly the same order of magnitude 
as $\tilde{m} \sim F/M_*$, where $F$ is the $F$-term VEV of the 
supersymmetry breaking field.  In this paper we do not discriminate 
the sizes of the two, and treat them to be of the same order:\ $m_{3/2} 
\sim \tilde{m}$.

\section{ISS with Intermediate Scale Colored Higgses}
\label{sec:model}

In this section, we present a simple model of ISS.  It is representative 
of a class of ISS GUTs where the only scale below the cutoff is that of 
supersymmetry breaking.  The model presented here differs from the one 
in Ref.~\cite{Hall:2013eko} in that the mass of the whole $H({\bf 5}) 
+ \bar{H}({\bf 5}^*)$ Higgs fields, of which the SM Higgs field is 
a part, now arises from supersymmetry breaking, where the numbers in 
parentheses denote representations under the $SU(5)$ GUT group.  This 
is achieved in a simple manner by obtaining the whole Higgs potential, 
including the one associated with the GUT-breaking field $\Sigma({\bf 24})$, 
from the K\"{a}hler potential.  We first describe the model and spectrum, 
discussing if/when the wino mass is lowered to the TeV scale due to a 
cancellation among various contributions as a result of environmental 
selection associated with the dark matter abundance.  We then discuss 
gauge coupling unification and proton decay.  We also present a detailed 
phenomenological analysis of the model in the case that the supersymmetry 
breaking parameters obey the mSUGRA-like boundary conditions at the 
supersymmetry breaking mediation scale $M_*$.

\subsection{Model}
\label{subsec:model}

We consider that physics below the cutoff scale $\Lambda \sim M_{\rm Pl}$ 
is well described by a supersymmetric GUT with the same field content as 
the minimal supersymmetric $SU(5)$ GUT.  The matter content of the model 
consists of the $\Sigma({\bf 24})$, $H({\bf 5})$, and $\bar{H}({\bf 5}^*)$ 
Higgs fields as well as three generations of the matter fields $T({\bf 10})$ 
and $\bar{F}({\bf 5}^*)$, where we have suppressed the generation indices. 
(Right-handed neutrino superfields $N({\bf 1})$'s will be introduced in 
Section~\ref{sec:scales} when we discuss neutrino masses.)

As in Ref.~\cite{Hall:2013eko}, we consider that the potential for the 
GUT-breaking field $\Sigma$ arises from the K\"{a}hler potential:
\begin{equation}
  K \supset \frac{c_2}{2} \Sigma^2 + \frac{c_3}{3\Lambda} \Sigma^3 
    + {\rm h.c.},
\label{eq:K-Sigma}
\end{equation}
where $c_{2,3}$ are dimensionless couplings of order unity, while $\Lambda 
\sim M_{\rm Pl}$ is the UV cutoff of the unified theory.  Similarly, here 
we also consider that the potential associated with the $H, \bar{H}$ Higgs 
fields arises from the K\"{a}hler potential terms
\begin{equation}
  K \supset d_2 H \bar{H} + \frac{d_3}{\Lambda} H \Sigma \bar{H} 
    + {\rm h.c.},
\label{eq:K-H}
\end{equation}
where $d_{2,3}$ are dimensionless couplings of order unity.  We assume 
that there is no interaction in the superpotential corresponding to the 
terms in Eqs.~(\ref{eq:K-Sigma},~\ref{eq:K-H}) at the fundamental level 
(i.e.\ in the supersymmetric flat-space limit).  This can be achieved, 
e.g., if the theory possesses an $R$ symmetry under which $\Sigma$ and 
$H\bar{H}$ are neutral:
\begin{equation}
  \Sigma(0), \qquad H\bar{H}(0).
\label{eq:R}
\end{equation}

When supersymmetry is broken and the cosmological constant is fine-tuned, 
the K\"{a}hler potential interactions of Eqs.~(\ref{eq:K-Sigma},~\ref{eq:K-H}) 
yield the effective superpotential through supergravity 
effects~\cite{Hall:1983iz}:%
\footnote{One way to see this is to use the conformal compensator 
 formalism~\cite{Siegel:1978mj}.  Using this formalism, the supergravity 
 Lagrangian in flat space is given by ${\cal L} \supset -3 M_{\rm Pl}^2 
 \int\!d^4\theta \Phi^\dagger \Phi\, e^{-K/3 M_{\rm Pl}^2}$, where 
 $\Phi = 1 + \theta^2 m_{3/2}$ is the conformal compensator field. 
 After canonically normalizing fields, $\Sigma \rightarrow \Sigma/\Phi$ 
 and similarly for $H$ and $\bar{H}$, this contains terms
 \begin{equation}
    {\cal L} \supset \int\!d^4\theta\, \Bigl( 
      \frac{c_2 \Phi^\dagger}{2 \Phi} \Sigma^2 
      + \frac{c_3 \Phi^\dagger}{3 \Lambda \Phi^2} \Sigma^3 
      + \frac{d_2 \Phi^\dagger}{\Phi} H \bar{H} 
      + \frac{d_3 \Phi^\dagger}{\Lambda \Phi^2} H \Sigma \bar{H} 
      + {\rm h.c.} \Bigr),
 \nonumber
 \end{equation}
 which, upon inserting $\Phi = 1 + \theta^2 m_{3/2}$, leads to 
 Eq.~(\ref{eq:W_eff}).}
\begin{equation}
  W_{\rm eff} = \frac{m_\Sigma}{2} \Sigma^2 
    + \frac{\lambda_\Sigma}{3} \Sigma^3 + \frac{m_H}{2} H \bar{H} 
    + \lambda_H H \Sigma \bar{H},
\label{eq:W_eff}
\end{equation}
where $m_\Sigma = c_2 m_{3/2}^*$, $\lambda_\Sigma = c_3 m_{3/2}^*/\Lambda$, 
$m_H = d_2 m_{3/2}^*$, and $\lambda_H = d_3 m_{3/2}^*/\Lambda$, so that
\begin{equation}
  m_\Sigma,\, m_H \sim O(\tilde{m}),
\qquad
  \lambda_\Sigma,\, \lambda_H \sim O\Bigl(\frac{\tilde{m}}{\Lambda}\Bigr).
\label{eq:orders}
\end{equation}
Here, we have used $m_{3/2} \sim \tilde{m}$.%
\footnote{If the supersymmetry breaking field $X$ is neutral, the K\"{a}hler 
 potential may contain terms of the form $X^\dagger \Sigma^2/\Lambda$, 
 $X^\dagger \Sigma^3/\Lambda^2$, $X^\dagger H \bar{H}/\Lambda$, and 
 $X^\dagger H \Sigma \bar{H}/\Lambda^2$.  They also lead to the effective 
 superpotential Eq.~(\ref{eq:W_eff}) with Eq.~(\ref{eq:orders}).}
The first two terms of Eq.~(\ref{eq:W_eff}) provide a non-zero VEV 
of $\Sigma$, $\langle \Sigma \rangle \sim m_\Sigma/\lambda_\Sigma \sim 
O(\Lambda)$, breaking $SU(5)$ to the SM gauge group.  In general, 
supersymmetry breaking effects in the $\Sigma$ potential lead to an 
$O(1)$ shift of the $\Sigma$ VEV,%
\footnote{In general, supersymmetry breaking effects, arising e.g.\ from 
 $X^\dagger X \Sigma^\dagger \Sigma/M_*^2$ and $X^\dagger X (\Sigma^\dagger 
 \Sigma)^2/M_*^2 \Lambda^2$, may give contributions to the $\Sigma$ 
 potential comparable to, or possibly larger than, the ones described 
 above.  We assume that these contributions do not eliminate a vacuum 
 breaking $SU(5)$ to the SM gauge group.}
giving a VEV for the $F$-component of $\Sigma$ of order $\tilde{m} \Lambda$:
\begin{equation}
  \langle \Sigma \rangle \sim O(\Lambda),
\qquad
  F_\Sigma \sim O(\tilde{m} \Lambda).
\label{eq:Sigma-VEVs}
\end{equation}
We take parameters of the model such that $\langle \Sigma \rangle$ is 
parametrically, e.g.\ a factor of a few to an order of magnitude, smaller 
than $\Lambda$, to ensure that there is an energy interval below $\Lambda$ 
in which physics is described by the $SU(5)$ theory.

Below the energy scale $\langle \Sigma \rangle$, which we will see is 
determined from gauge coupling unification as
\begin{equation}
  \langle \Sigma \rangle \sim 10^{16}\mbox{--}10^{17}~{\rm GeV},
\label{eq:VEV-Sigma}
\end{equation}
the massive vector supermultiplets containing the GUT gauge bosons decouple, 
and physics is well described by the low-energy supersymmetric gauge 
theory with the SM gauge group $SU(3)_{\rm C} \times SU(2)_{\rm L} 
\times U(1)_{\rm Y}$.  An important property of the superpotential 
in Eq.~(\ref{eq:W_eff}) with Eq.~(\ref{eq:orders}) is that because of 
the overall $\tilde{m}$ factor, the masses of the uneaten components of 
$\Sigma$---$\Sigma_8({\bf 8}, {\bf 1})_0$, $\Sigma_3({\bf 1}, {\bf 3})_0$, 
and $\Sigma_1({\bf 1}, {\bf 1})_0$ where the numbers represent the SM 
gauge quantum numbers---are at the intermediate scale:
\begin{equation}
  M_{\Sigma_8} \sim M_{\Sigma_3} \sim M_{\Sigma_1} \sim O(\tilde{m}),
\label{eq:intermed-Sigma}
\end{equation}
where superpartners of the Supersymmetric Standard Model (SSM) 
exist~\cite{Hall:2013eko}.  This potentially has cosmological implications. 
For example, if the Hubble parameter during inflation is large $H_I > 
\tilde{m}$, as suggested by the recent BICEP2 observation~\cite{Ade:2014xna}, 
then depending on the dynamics during inflation (e.g.\ the sign of 
the Hubble induced mass-squared for $\Sigma$), the GUT symmetry may 
be recovered during inflation, which would lead to unwanted monopole 
production after inflation.  Below, we assume that such symmetry 
recovery does not occur.

\begin{figure}[t!]
\begin{center}
  {\includegraphics[clip]{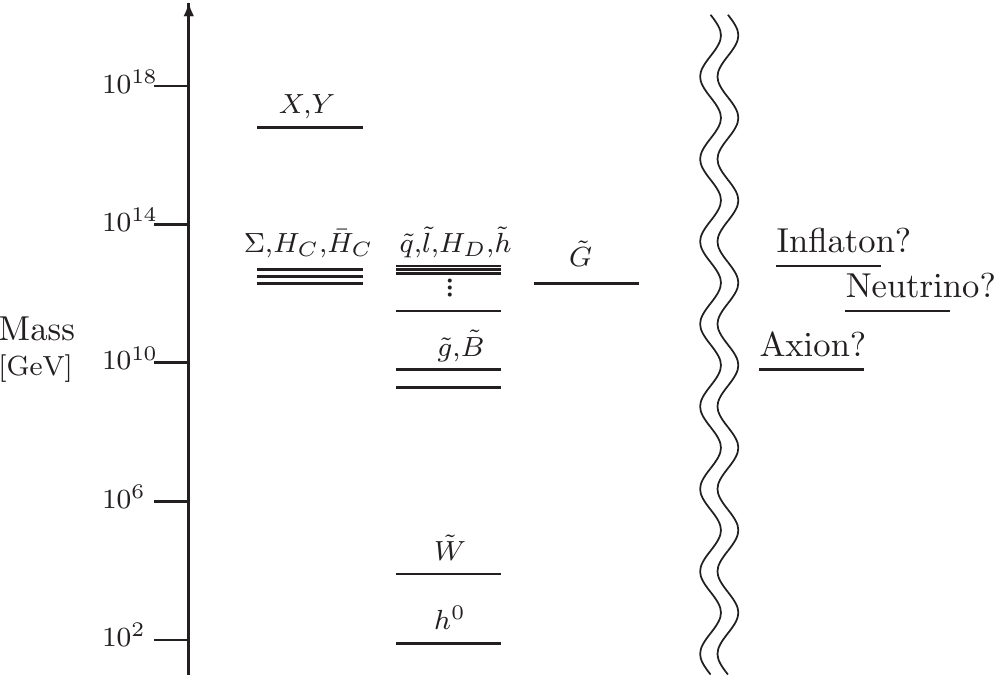}}
\caption{A schematic depiction of the spectrum of the model.  Here, 
 $X,Y$ represent the massive GUT gauge supermultiplets, $\Sigma$ uneaten 
 components of the $\Sigma({\bf 24})$ superfield, and $H_C$ and $\bar{H}_C$ 
 the colored Higgs supermultiplets; the other symbols denote particles 
 in the SSM in the self-explanatory notation (with $\tilde{G}$ being the 
 gravitino).  We have depicted the wino, $\tilde{W}$, at a TeV scale, 
 although it may be at the intermediate scale $\tilde{m}$ if the reheating 
 temperature after inflation is sufficiently low or $R$-parity is broken; 
 see discussions in Section~\ref{subsec:wino}.  The scale $\tilde{m}$ 
 can lead to neutrino masses, inflation, and axions, as indicated on 
 the right.}
\label{fig:spectrum}
\end{center}
\end{figure}

In a similar manner, in the present model the masses of the 
colored Higgs fields, as well as those of the second Higgs doublet 
$H_D$ and Higgsino $\tilde{h}$ of the SSM, are also at the intermediate 
scale.  In the minimal $SU(5)$ GUT, the doublet Higgses $H_u$ and $H_d$ 
of the SSM are embedded in the fundamental representations of $SU(5)$ 
as $H = (H_u, H_C)$ and $\bar{H} = (H_d, \bar{H}_C)$.  The superpotential 
of Eq.~(\ref{eq:W_eff}) implies that the masses for all these fields are 
also of order $\tilde{m}$:
\begin{equation}
  M_{H_C} \sim M_{H_D} \sim M_{\tilde{h}} \sim O(\tilde{m}),
\label{eq:m-H_C}
\end{equation}
except for the light Higgs doublet $h$ of the SM, which is environmentally 
selected to be of order the weak scale $v \ll \tilde{m}$.  One might 
think that such low values for the colored Higgs masses are excluded 
by the proton decay constraints.  This is, however, not the case as will 
be discussed in Section~\ref{subsec:p-decay}.  In Fig.~\ref{fig:spectrum}, 
we give a schematic depiction of the spectrum of the present model. 
In the figure, we have put the wino, $\tilde{W}$, to be at a TeV scale, 
since its mass may be environmentally selected to be in this range; see 
Section~\ref{subsec:wino} below.

There are several virtues in the model presented here, with the last 
two terms in Eq.~(\ref{eq:W_eff}) arising from the K\"{a}hler potential, 
compared to the model in Ref.~\cite{Hall:2013eko}, in which these terms 
exist in the superpotential before supersymmetry breaking with $m_H \sim 
O(\langle \Sigma \rangle)$ and $\lambda_H \sim O(1)$.  First, since 
the supersymmetric masses of $\Sigma_3$ and $\Sigma_1$ are comparable 
to $\tilde{m}$, the interaction $\lambda_H H \Sigma \bar{H}$ in the 
superpotential gives a non-decoupling contribution to the Higgs quartic 
coupling:\ $\delta \lambda \sim \lambda_H^2 \sin^2 2\beta$.  In order 
to preserve the identification of $\tilde{m}$ to be at a scale close to 
the point in which $\lambda$ crosses zero, as in Eq.~(\ref{eq:m-tilde}), 
this contribution needs to be small, $\lambda_H \lesssim 0.1$.  In 
the model of Ref.~\cite{Hall:2013eko}, this condition needs to be 
imposed by hand, while here it is automatic because $\lambda_H \sim 
O(\tilde{m}/\Lambda) \ll 1$.  Note that since $m_H \sim O(\tilde{m})$ 
and $\lambda_H \sim O(\tilde{m}/\Lambda)$, the present model does not 
require doublet-triplet splitting (except for the fine-tuning needed 
to make the SM Higgs light); namely, the contributions to the mass 
of the heavy Higgs doublet from the third and fourth terms in 
Eq.~(\ref{eq:W_eff}) need not be nearly canceled with each other. 
This allows us to have a natural size of $F_\Sigma \sim O(\tilde{m} 
\Lambda)$ in Eq.~(\ref{eq:Sigma-VEVs}) while still allowing for 
successful electroweak symmetry breaking, since it only leads to 
the holomorphic supersymmetry-breaking mass for the Higgs doublets 
of order $\mu B \approx \lambda_H F_\Sigma \sim O(\tilde{m}^2)$. 
(In the model of Ref.~\cite{Hall:2013eko}, $F_\Sigma \sim O(\tilde{m} 
\Lambda)$ leads to a too large $\mu B$ term of order $\tilde{m} 
\Lambda$, so that $F_\Sigma$ must be suppressed by extra environmental 
selection.)  Finally, the fact that $m_H \sim O(\tilde{m})$ implies 
that the level of fine-tuning needed to reproduce electroweak symmetry 
breaking is of order $v^2/\tilde{m}^2$ in the present model, while 
the one in Ref.~\cite{Hall:2013eko} requires an extra factor of order 
$\tilde{m}^2/\Lambda^2$ to keep $\mu^2$ and $\mu B$ to be of order 
$\tilde{m}^2$.  While none of the above issues excludes the model in 
Ref.~\cite{Hall:2013eko}, their absence adds an aesthetic appeal to 
the model discussed here.

The electroweak symmetry is broken by the VEV of the light SM Higgs 
doublet $h$, whose mass-squared parameter (and thus VEV) is fine-tuned 
to be of order the weak scale due to environmental selection.  Specifically, 
the mass-squared matrix for the two Higgs doublets at the scale $\tilde{m}$ 
is given by
\begin{equation}
  {\cal M}_{\rm H}^2 = \begin{pmatrix}
    |\mu|^2 + m_{H_u}^2 & \mu B \\
    \mu B & |\mu|^2 + m_{H_d}^2
 \end{pmatrix},
\label{eq:MH2}
\end{equation}
where $|\mu|^2$, $|\mu B|$, $|m_{H_u}^2|$, $|m_{H_d}^2|$ are of 
order $\tilde{m}^2$.  These parameters are chosen such that one Higgs 
doublet remain below $\tilde{m}$, i.e.\ the determinant of the matrix 
${\cal M}_{\rm H}^2$ to be extremely small compared with its natural size 
$\sim \tilde{m}^4$.  The resulting SM Higgs doublet is given by the 
combination $h \approx \sin\beta\, h_u + \cos\beta\, h_d^\dagger$ with
\begin{equation}
  \tan^2\!\beta = \frac{|\mu|^2 + m_{H_d}^2}{|\mu|^2 + m_{H_u}^2}.
\label{eq:tan-beta}
\end{equation}
Since the quartic coupling for the Higgs is given by
\begin{equation}
  \lambda(\tilde{m}) = \frac{g^2 + g'^2}{4}\, \cos^2\!2\beta,
\label{eq:lambda}
\end{equation}
where $g$ and $g'$ are the $SU(2)_{\rm L}$ and $U(1)_{\rm Y}$ gauge 
couplings at $\tilde{m}$, we consider $\tan\beta \sim 1$.  Such values of 
$\tan\beta$ are easily obtained, e.g., if $|\mu|^2 \gtrsim |m_{H_u,H_d}^2|$ 
or if $m_{H_u}^2$ and $m_{H_d}^2$ are nearly degenerate; see also the 
discussion in Section~\ref{subsec:mSUGRA}.  With electroweak symmetry 
breaking, the SM quarks and leptons obtain masses through the standard 
Yukawa interactions in the superpotential
\begin{equation}
  W = y_U TTH + y_D T \bar{F} \bar{H} 
  + \frac{\eta_U}{\Lambda} \Sigma T T H 
  + \frac{\eta_D}{\Lambda} \Sigma T \bar{F} \bar{H} + \cdots,
\label{eq:Yukawa}
\end{equation}
where we have suppressed the generation indices.  Higher-dimension terms 
involving $\Sigma$ are needed to correct unwanted $SU(5)$ relations for 
the quark/lepton masses.

To summarize, the model is characterized by the following holomorphic 
part of the K\"{a}hler potential, $\tilde{K}$, and superpotential, $W$:
\begin{align}
  \tilde{K} &= \Lambda^2\, f\Bigl( \frac{\Sigma}{\Lambda}, 
    \frac{H\bar{H}}{\Lambda^2} \Bigr),
\label{eq:model-K}\\
  W &= y_U\Bigl( \frac{\Sigma}{\Lambda} \Bigr)\, T T H 
    + y_D\Bigl( \frac{\Sigma}{\Lambda} \Bigr)\, T \bar{F} \bar{H}, 
\label{eq:model-W}
\end{align}
(except for the terms needed for neutrino masses; see 
Section~\ref{subsec:neutrino}), where $f$ is a holomorphic function 
with the coefficients expected to be of order unity, and $y_U$ 
and $y_D$ are holomorphic functions associated with the Yukawa 
couplings. Here, we have assumed $R$-parity.  The form of 
Eqs.~(\ref{eq:model-K},~\ref{eq:model-W}), including $R$ parity 
conservation, can be easily enforced by an $R$ symmetry; for 
example, we may assign a neutral $R$ charge to $\Sigma$, $H$, 
and $\bar{H}$, as in Eq.~(\ref{eq:R}), and a unit charge to $T$ 
and $\bar{F}$. (A different charge assignment will be considered 
in Section~\ref{subsec:neutrino}.)  All the dimensionful 
parameters, except $\Lambda$, are generated through supersymmetry 
breaking $\tilde{m}$, leading to the effective superpotential of 
Eq.~(\ref{eq:W_eff}).  The GUT symmetry is broken at $\langle \Sigma 
\rangle \sim O(\Lambda)$, while the electroweak symmetry---due to 
environmental selection---at $\langle h \rangle = v \ll \tilde{m}$.

We finally discuss the gaugino masses.  Unlike scalar superpartners, 
the gaugino masses may be protected against supersymmetry breaking 
effects via some symmetry.  For example, if the supersymmetry breaking 
field $X$ has some charge, its direct coupling to the gauge field-strength 
superfields $[X {\cal W}^\alpha {\cal W}_\alpha /\Lambda]_{\theta^2}$ 
is suppressed.  There are, however, many other sources for the gaugino 
masses:\ anomaly mediation~\cite{Randall:1998uk}, threshold corrections 
from $H$ and/or $\Sigma$, and the higher dimension operator $[\Sigma 
{\cal W}^\alpha {\cal W}_\alpha /\Lambda]_{\theta^2}$ with $F_\Sigma 
\neq 0$.  In particular, since the operator $[\Sigma {\cal W}^\alpha 
{\cal W}_\alpha /\Lambda]_{\theta^2}$ is used to reproduce the observed 
SM gauge couplings (see Section~\ref{subsec:gcu}) and we naturally 
expect $F_\Sigma \sim O(\tilde{m} \Lambda)$ (see Eq.~(\ref{eq:Sigma-VEVs})), 
the last contribution gives typically the gaugino masses not much smaller 
than $\tilde{m}$.  As we will see in the next subsection, however, the 
wino mass may have to be lowered to a TeV scale as a result of environmental 
selection associated with dark matter.  This occurs through a cancellation 
of various contributions given above, which in turn could suppress the 
gluino and bino masses through GUT relations.  Note that the cancellation 
of the wino mass requires a modest suppression of $F_\Sigma$ and/or the 
coefficient of $[\Sigma {\cal W}^\alpha {\cal W}_\alpha /\Lambda]_{\theta^2}$ 
to allow the cancellation of its contribution with the rest, which are 
one-loop suppressed.  We thus expect that the gaugino masses are in 
the range
\begin{equation}
  M_\lambda \approx O(10^{-2}\mbox{--}1)\, \tilde{m},
\label{eq:gaugino}
\end{equation}
except possibly for the wino, which may be at a TeV scale.

\subsection{TeV-scale wino}
\label{subsec:wino}

If $R$-parity is unbroken, the lightest supersymmetric particle (LSP) is 
stable and contributes to the dark matter once cosmologically produced.%
\footnote{In the present model, $R$-parity may naturally arise 
 as a ${\bf Z}_2$ subgroup of the $U(1)_R$ symmetry described in 
 Eq.~(\ref{eq:R}).  For example, for the $R$ charge assignment considered 
 below Eqs.~(\ref{eq:model-K},~\ref{eq:model-W}), $R$-parity arises as 
 an unbroken ${\bf Z}_2$ subgroup of $U(1)_R$ after supersymmetry breaking, 
 more precisely, after a constant term in the superpotential is introduced 
 to cancel the cosmological constant induced by supersymmetry breaking.}
The abundance of the LSP in the universe may depend strongly on the 
reheating temperature $T_R$ after inflation as well as the branching 
ratio of the inflaton decay into the LSP.  Here we see that this most 
likely requires the mass of the LSP, $m_{\rm LSP}$, to be in the TeV 
region.  Such a small LSP mass may result from a cancellation of various 
contributions as a result of environmental selection associated with 
dark matter~\cite{Hall:2011jd}.

Let us first consider the case in which $T_R \gtrsim m_{\rm LSP}$.  In 
this case, the LSP is thermalized and its abundance is roughly given by
\begin{equation}
  \Omega^{\rm th} h^2 \sim 10^{16} 
    \left( \frac{m_{\rm LSP}}{10^{12}~{\rm GeV}} \right)^2.
\label{eq:Omega-th}
\end{equation}
This grossly overcloses the universe for $m_{\rm LSP} \sim O(\tilde{m})$. 
We now consider the case $T_R \ll m_{\rm LSP}$.  In this case, thermal 
gas scattering and inflaton decay may still lead to a significant 
amount of the LSP abundance.  From thermal scattering, we obtain the 
approximate LSP abundance of
\begin{equation}
  \Omega^{\rm sc} h^2 \sim 10^{21} 
    \left( \frac{T_R}{m_{\rm LSP}} \right)^7.
\label{eq:Omega-sc}
\end{equation}
(A similar estimate in a different context can be found in 
Ref.~\cite{Takayama:2007du}.)  Furthermore, if the mass of the inflaton 
$m_\phi$ is sufficiently larger than $m_{\rm LSP}$, the inflaton may 
directly decay into LSPs which do not effectively annihilate afterwards. 
The resulting LSP abundance is then roughly given by
\begin{equation}
  \Omega^{\rm dec} h^2 \sim 10^{20}\, B_\phi 
    \left( \frac{T_R}{m_\phi} \right) 
    \left( \frac{m_{\rm LSP}}{10^{12}~{\rm GeV}} \right),
\label{eq:Omega-dec}
\end{equation}
where $B_\phi$ is the branching fraction of the inflaton decay to 
the LSP.  We thus find that unless $T_R \lesssim 10^{-3}\, m_{\rm LSP}$ 
{\it and} $B_\phi$ is essentially zero, the LSP with $m_{\rm LSP} \sim 
O(\tilde{m})$ will overclose the universe.

The mass of the LSP, however, may be environmentally selected:\ it 
may be reduced to the TeV region due to a cancellation of various 
contributions~\cite{Hall:2011jd}.  This occurs if there are environmental 
constraints that strongly disfavor observers in universes with much more 
dark matter than our own, as argued, e.g., in Ref.~\cite{Tegmark:2005dy}. 
Here and below, we assume that the LSP is the wino, $\tilde{W}$. 
In this case, if the wino mass is smaller than about $3~{\rm TeV}$, 
$\Omega^{\rm th} h^2 < 0.1$~\cite{Hisano:2006nn}.  In general, the 
selection effects for dark matter act on any candidate, no matter 
the production mechanism, so dark matter may be multi-component; 
in particular, axions may comprise a part of the dark matter.  This 
consideration, therefore, gives the only upper bound on the wino mass:\ 
$M_{\tilde{W}} \lesssim 3~{\rm TeV}$.

An important signal for a TeV-scale wino is direct production at colliders. 
The charged wino is slightly heavier than the neutral wino by $\simeq 
165~{\rm MeV}$~\cite{Feng:1999fu}.  The small mass difference makes the 
charged wino live long:\ $c\tau = \mbox{a few}~{\rm cm}$, which can be 
detected as a disappearing track at the LHC~\cite{Ibe:2006de,Asai:2007sw}. 
The current LHC bound for the wino mass is $m_{\tilde{W}} > 270~{\rm GeV}$ 
at 95\% confidence level~\cite{Aad:2013yna}.  At a future lepton collider, 
direct observation of such a charged track is important.  In addition 
to direct production, processes mediated by wino loops may also provide 
clues for wino search; see e.g.\ Ref.~\cite{KEK-talk}.

Another promising way of searching for a TeV-scale wino is through 
cosmic-ray signals from wino dark matter annihilation.  The annihilation 
of winos leads to a variety of cosmic-ray species, e.g.\ line and continuum 
photons~\cite{Hisano:2004ds} and antiprotons~\cite{Hisano:2005ec}, 
whose cross section may be enhanced by the Sommerfeld effect.  Recent 
observations of $\gamma$-rays by the H.E.S.S.\ and Fermi experiments 
give important constraints, although they are subject to astrophysical 
uncertainties~\cite{Cohen:2013ama,Hryczuk:2014hpa}.  Cosmic-ray antiprotons 
can also provide a powerful tool for searching for wino dark matter. 
While this signal also suffers from astrophysical uncertainties, the 
on-going AMS-02 experiment can reduce such uncertainties~\cite{Pato:2010ih}, 
so that this may allow us to probe essentially all the wino mass range 
if it is the dominant component of the dark matter~\cite{Hall:2012zp}.

In summary, unless $T_R \lesssim 10^{-3}\, m_{\rm LSP}$ and $B_\phi 
\approx 0$ (or $R$-parity is broken), the mass of the LSP must be much 
smaller than $\tilde{m}$, and for the wino LSP
\begin{equation}
  270~{\rm GeV} < M_{\tilde{W}} \lesssim 3~{\rm TeV}.
\label{eq:m_wino}
\end{equation}
(In addition, small portions of this mass range may be excluded by dark 
matter constraints; see e.g.\ Ref.~\cite{Hryczuk:2014hpa}.)  The spectrum 
of the model in this case is depicted in Fig.~\ref{fig:spectrum}.  Below, 
we consider both this TeV-scale wino case and the case with $m_{\rm LSP} 
\sim O(\tilde{m})$.

\subsection{Gauge coupling unification}
\label{subsec:gcu}

We now discuss unification of the SM gauge couplings in the ISS model 
described here.  Following Ref.~\cite{Hall:2013eko}, we consider two 
variables
\begin{align}
  R_X &= \frac{1}{\sqrt{38}} 
    \left( \frac{5}{g_1^2} -\frac{3}{g_2^2} - \frac{2}{g_3^2} \right),
\label{eq:R_X}\\
  R_H &= \frac{1}{\sqrt{14}} 
    \left( \frac{3}{g_2^2} -\frac{2}{g_3^2} - \frac{1}{g_1^2} \right).
\label{eq:R_H}
\end{align}
In the absence of higher-dimensional gauge kinetic operators involving 
$\Sigma$, the energies at which $R_X$ and $R_H$ cross zero would 
correspond to the masses of the $XY$ GUT gauge fields, $M_X$, and 
the colored Higgs fields, $M_{H_C}$, respectively.  In general, however, 
we expect the gauge kinetic function contains higher-dimensional terms 
involving $\Sigma$:
\begin{equation}
  {\cal L} \supset \frac{1}{2 g^2} \int\! d^2\theta\, 
      \bigg\{ {\rm Tr}[{\cal W}^\alpha {\cal W}_\alpha] 
    + \frac{a}{\Lambda} {\rm Tr}[\Sigma {\cal W}^\alpha {\cal W}_\alpha] 
    + O(\Sigma^2) \biggr\} + {\rm h.c.},
\label{eq:threshold}
\end{equation}
giving GUT-breaking threshold corrections with $\langle \Sigma \rangle 
\neq 0$, where $g$ and ${\cal W}^\alpha$ are the $SU(5)$ gauge coupling 
and gauge field-strength superfield, respectively.  An important point 
is that the leading dimension-five operator (the second term above) gives 
a correction to $R_H$, but not to $R_X$---$R_X$ is corrected only at 
order $\langle \Sigma \rangle^2/\Lambda^2$, which is small.  We can, 
therefore, read off the mass of the $XY$ gauge boson, $M_X \approx 
\langle \Sigma \rangle$, by plotting $R_X$ as a function of energy and 
seeing where it crosses zero,
\begin{equation}
  R_X(M_X) \approx 0.
\label{eq:R_X-M_X}
\end{equation}
On the other hand, since $R_H$ receives a relatively large correction from 
the dimension-five operator, it does not strongly constrain $M_{H_C}$---any 
value of $M_{H_C}$ is consistent as long as $R_H$ at that scale is reasonably 
small
\begin{equation}
  \left| R_H(M_{H_C}) \right| 
  \approx \left| -\frac{a V}{\Lambda} \right| 
  \lesssim O(0.1),
\label{eq:R_H-M_H}
\end{equation}
where $V$ is the GUT breaking VEV, $\langle \Sigma \rangle = V\, 
{\rm diag}(2,2,2,-3,-3)/\sqrt{60}$.

\begin{figure}[t!]
\begin{center}
\subfigure[$M_\lambda = m_0 = m_\Sigma = m_{H_C} = 10^{12}~{\rm GeV}$]
  {\includegraphics[clip, width = 0.48 \textwidth]{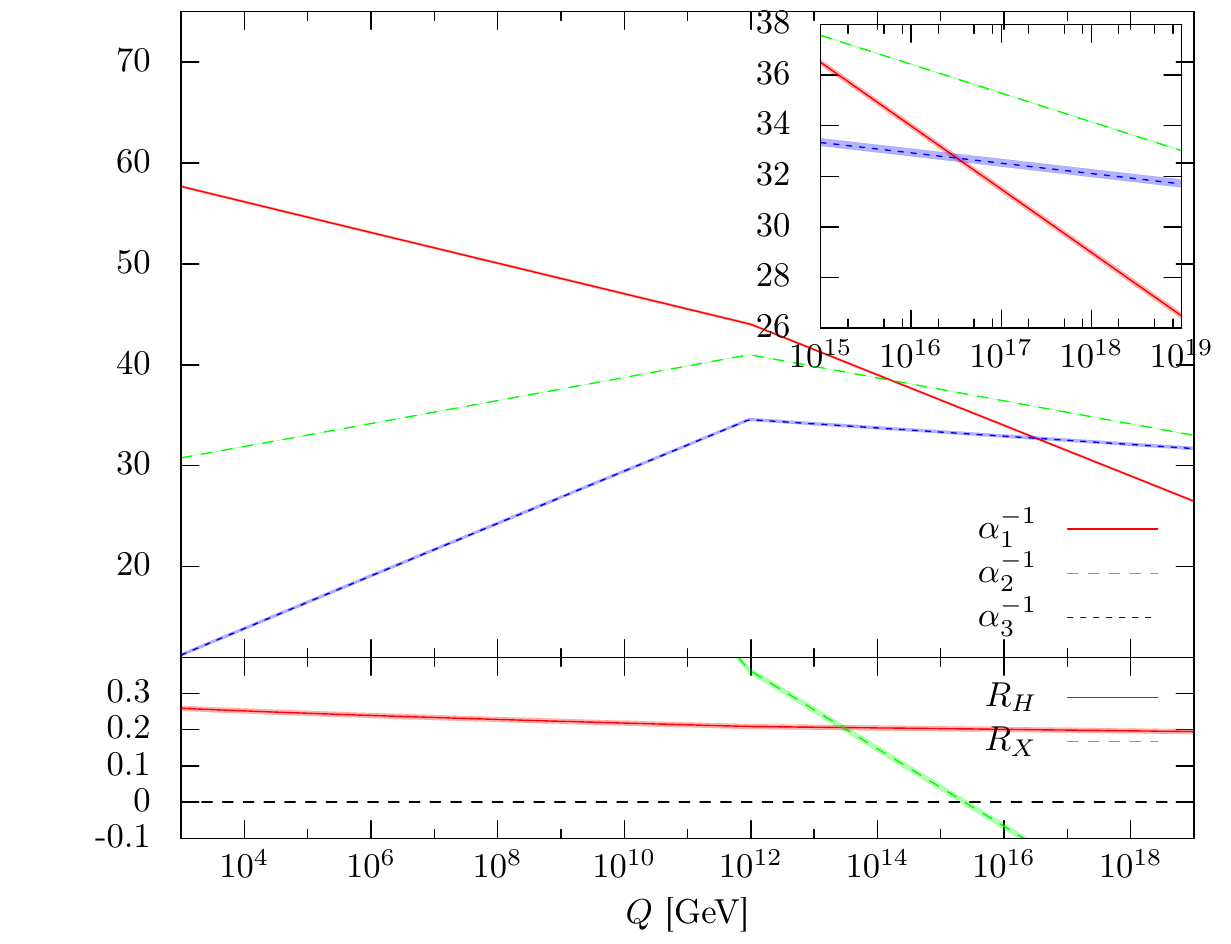}
  \label{fig:gauge:a}}
\subfigure[$10^{2} M_\lambda = m_0 = m_\Sigma =  m_{H_C} = 10^{12}~{\rm GeV}$]
  {\includegraphics[clip, width = 0.48 \textwidth]{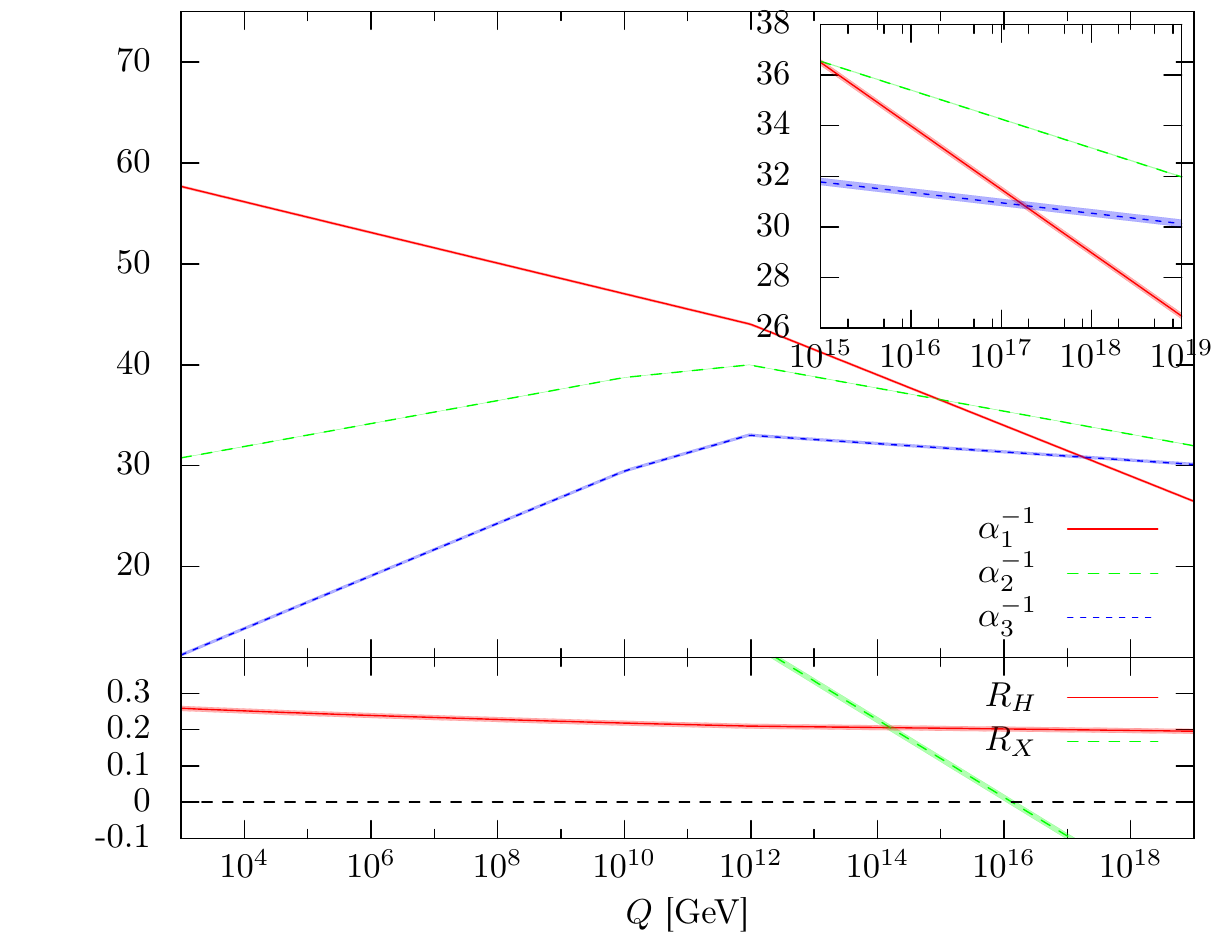}
  \label{fig:gauge:b}}
\subfigure[Same as (a) except that $M_{\tilde{W}} = 1~{\rm TeV}$]
  {\includegraphics[clip, width = 0.48 \textwidth]{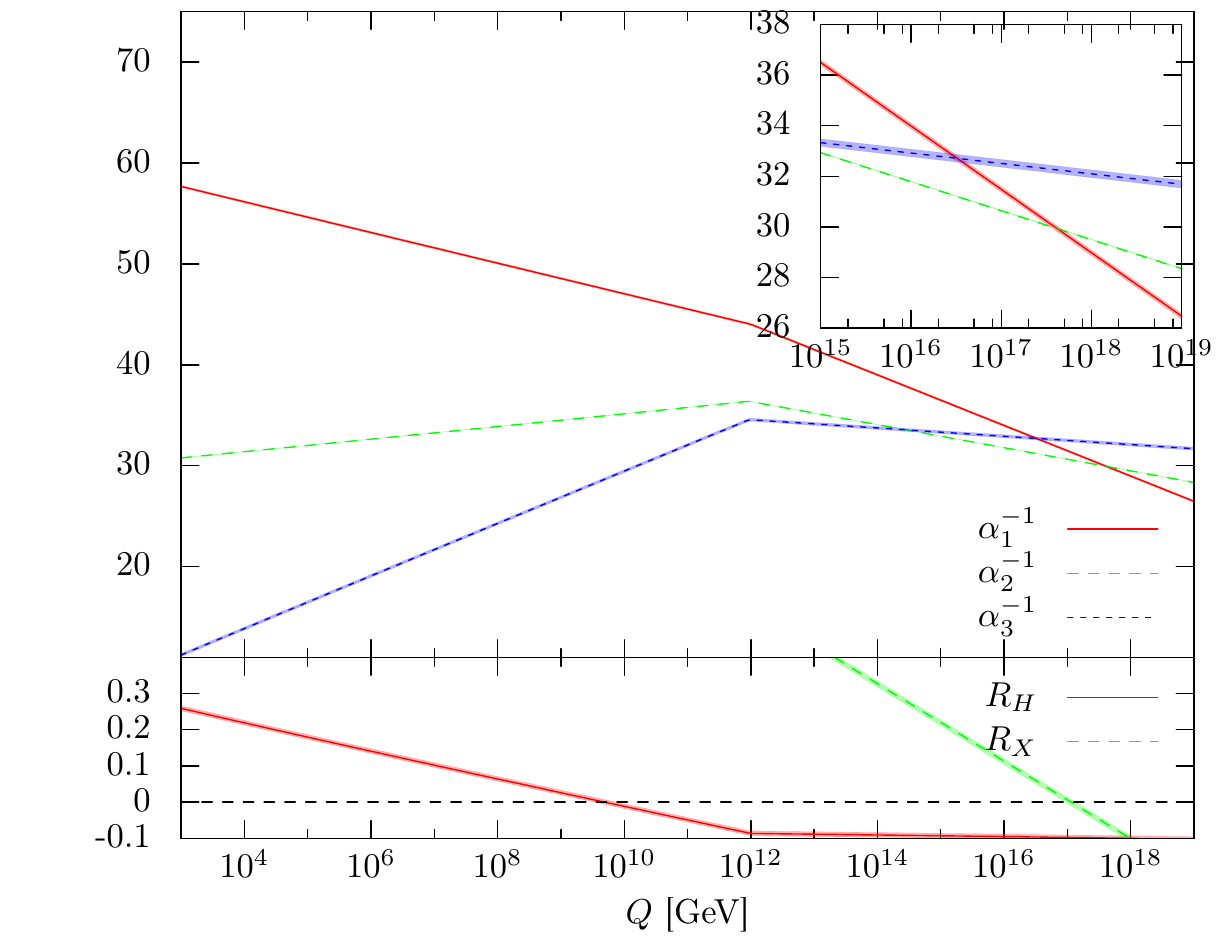}
  \label{fig:gauge:c}}
\subfigure[Same as (b) except that $M_{\tilde{W}} = 1~{\rm TeV}$]
  {\includegraphics[clip, width = 0.48 \textwidth]{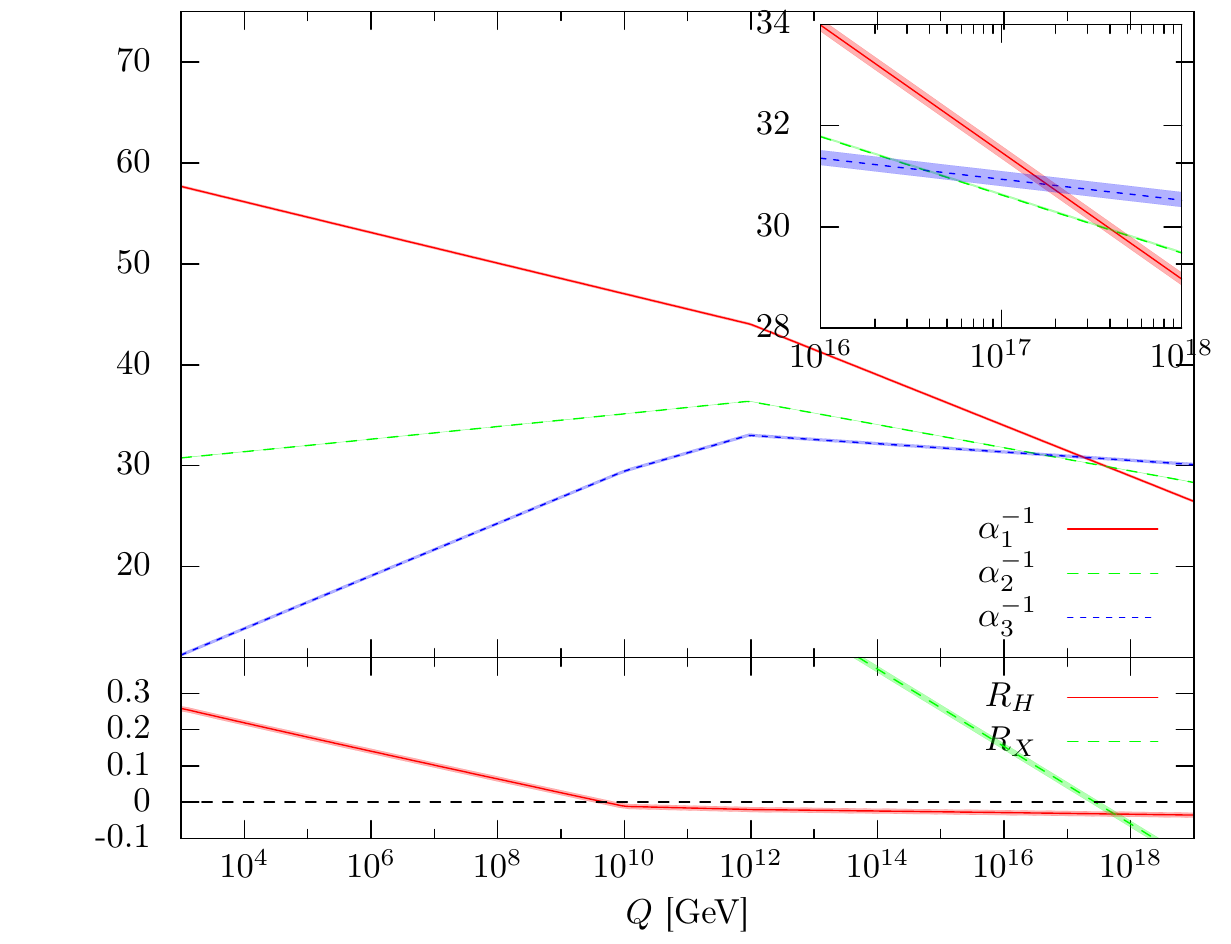}
  \label{fig:gauge:d}}
\caption{The renormalization group running of the SM gauge couplings for 
 representative spectra in the ISS model.  Each shaded band represents 
 the 3-$\sigma$ experimental uncertainty for the gauge coupling.  Here, 
 we have assumed $\tan\beta = 1$.  Important combinations of the gauge 
 couplings, $R_X$ and $R_H$ defined in Eqs.~(\ref{eq:R_X},~\ref{eq:R_H}), 
 are also plotted; they determine the scale and precision of unification, 
 as in Eqs.~(\ref{eq:R_X-M_X},~\ref{eq:R_H-M_H}).}
\label{fig:gauge}
\end{center}
\end{figure}
In Fig.~\ref{fig:gauge}, we show the running of the SM gauge couplings 
for some selected ISS mass spectra.  In Fig.~\ref{fig:gauge:a} we show 
the case in which all the superpartners and (uneaten) GUT particles have 
a common mass of $\tilde{m} = 10^{12}~{\rm GeV}$ (corresponding to the 
case with $m_{\rm LSP} \sim O(\tilde{m})$ in the previous subsection, 
while in Fig.~\ref{fig:gauge:b} we set the gaugino masses $M_\lambda$ 
to be suppressed by a factor of $100$ compared with the rest of the 
intermediate scale particles.  We find that the unification scale, 
determined by Eq.~(\ref{eq:R_X-M_X}), is
\begin{equation}
  M_X \sim 10^{16}~{\rm GeV}
  \quad\mbox{for}\quad m_{\rm LSP} \sim O(\tilde{m}).
\label{eq:MX_case-1}
\end{equation}
The size of the threshold correction, determined by Eq.~(\ref{eq:R_H-M_H}) 
with $M_{H_C} \sim O(\tilde{m})$, is $|aV/\Lambda| \approx 0.2$.  In 
Figs.~\ref{fig:gauge:c} and \ref{fig:gauge:d}, we show the same plots 
as Figs.~\ref{fig:gauge:a} and \ref{fig:gauge:b}, respectively, except 
that the wino mass is lowered to $1~{\rm TeV}$.  This slightly raises 
the unification scale
\begin{equation}
  M_X \sim 10^{17}~{\rm GeV}
  \quad\mbox{for}\quad m_{\tilde{W}} \sim {\rm TeV},
\label{eq:MX_case-2}
\end{equation}
and improves the precision for gauge coupling unification; the required 
size of the threshold correction from the dimension-five operator is now 
$|aV/\Lambda| \lesssim 0.1$.

We finally comment on bottom-tau Yukawa unification.  In the minimal ISS 
model discussed here, the ratio of the two couplings is $y_b/y_\tau \simeq 
0.6$ at the GUT scale, so that it requires a relatively large threshold 
correction to be compatible with the GUT embedding of the quarks and 
leptons.  This can be achieved, for example, by taking $(y_D)_{33} 
\lesssim (\eta_D)_{33}$ in Eq.~(\ref{eq:Yukawa}).  Similar GUT-violating 
contributions may also be needed to accommodate the observed quark and 
lepton masses for lighter generations.

\subsection{Proton decay}
\label{subsec:p-decay}

Here we discuss proton decay.  As we have seen, the mass of the $XY$ 
GUT gauge bosons, $M_X$, is comparable or larger than in the conventional 
weak-scale SSM.  In particular, when the wino is at a TeV, $M_X \sim 
10^{17}~{\rm GeV}$ as in Eq.~(\ref{eq:MX_case-2}), so that dimension-six 
proton decay caused by gauge boson exchange is suppressed.

How about proton decay caused by exchange of colored Higgs fields, which 
now have masses of order $\tilde{m} \ll \langle \Sigma \rangle$?  In 
the conventional weak-scale SSM, the colored Higgs supermultiplets 
$H_C$ and $\bar{H}_C$ induce large proton decay rates.  In this case 
the dominant contributions come from one-loop diagrams involving 
weak-scale superpartners with amplitudes suppressed only by 
$1/(M_{H_C} m_{\rm soft})$, where $m_{\rm soft} \sim v$ is the mass 
of the weak-scale superpartners.  To avoid rapid proton decay, we 
need to push the mass of the colored Higgs multiplets to be very 
large~\cite{Murayama:2001ur}.  If the sfermion mass scale is much larger 
than the weak scale, however, the proton decay rate from these processes 
(dimension-five proton decay) can be suppressed, and the constraints 
can accordingly be relaxed~\cite{Hisano:2013exa}.

\begin{figure}[t!]
\begin{center}
  \subfigure[Tree level]
    {\includegraphics[clip, width = 0.2 \textwidth]{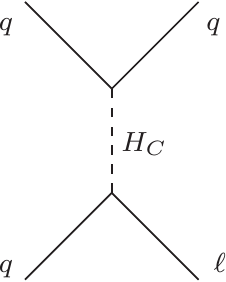}
    \label{fig:diagram:tree}}
\hspace{0.05 \textwidth}
  \subfigure[One-loop level]
    {\includegraphics[clip, width = 0.2 \textwidth]{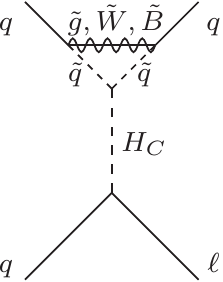}
    \label{fig:diagram:loop}}
\hspace{0.05 \textwidth}
  \subfigure[Correction to Yukawa couplings]
    {\includegraphics[clip, width = 0.4 \textwidth]{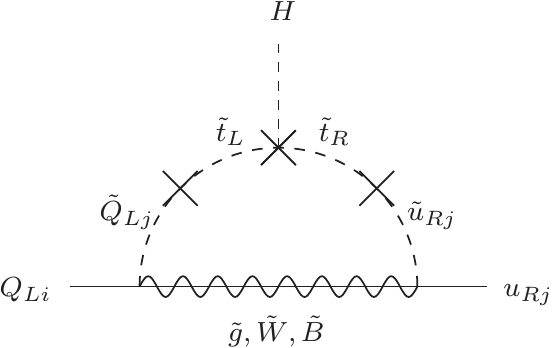}
    \label{fig:diagram:yukawa}}
\caption{Examples of diagrams relevant for proton decay.}
\label{fig:diagram}
\end{center}
\end{figure}

In ISS models, the sfermion mass scale is quite large, $\tilde{m} \gg v$, 
so that dimension-five proton decay can be suppressed, which allows us 
to take $M_{H_C} \sim O(\tilde{m})$ as has been described so far.  In 
fact, unlike the conventional case, the dominant contribution to proton 
decay typically comes from tree-level colored Higgs-boson exchange, as 
shown in Fig.~\ref{fig:diagram:tree}.  This contribution is suppressed 
by $1/M_{H_C}^2$ in the amplitude, and is negligible in conventional GUTs; 
but here the suppression is only $1/M_{H_C}^2 \sim O(1/\tilde{m}^2)$, 
and its amplitude is larger than that for dimension-five proton decay 
by a one-loop factor.  The dominant decay mode is expected to be $p 
\rightarrow \bar{\nu} K^+$, with lifetime given approximately by
\begin{equation}
  \tau_p \approx O(10^{32}\mbox{--}10^{33}) 
    \left( \frac{M_{H_C}}{10^{11}~{\rm GeV}} \right)^4~{\rm years}.
\label{eq:tau_p}
\end{equation}
The current limit from the Super-Kamiokande experiment is 
$\tau_p(p \rightarrow \bar{\nu} K^+) > 5.9 \times 10^{33}~{\rm years}$ 
at 90\% confidence level~\cite{SK}, so that $M_{H_C}$ greater than 
$O(10^{11}~{\rm GeV})$ is consistent with the latest observation. 
This limit is expected to be improved to $2.5 \times 10^{34}~{\rm year}$ 
in the hyper-Kamiokande experiment~\cite{Abe:2011ts}.

The proton decay rate in Eq.~(\ref{eq:tau_p}) is subject to several 
uncertainties.  One of them comes from GUT $CP$ phases; there are two 
additional physical $CP$ phases in the colored Higgs Yukawa couplings, 
which cannot be determined by the SSM Yukawa couplings.  Depending on 
these phases, cancellation between Wilson operators causing the proton 
decay may occur.  This leads to an $O(10)$ uncertainty in the estimate 
of the proton lifetime.  (We will see this uncertainty explicitly in 
the study of the mSUGRA example in the next subsection.)  Another source 
of uncertainties comes from the long distance QCD contribution to the 
proton decay matrix elements, which also leads to an $O(10)$ uncertainty 
in the lifetime estimate.  Furthermore, as we discussed before, 
accommodating the observed quark and lepton masses in the model 
requires contributions from higher-dimensional operators to the 
Yukawa couplings.  These operators also affect the estimate of 
the proton decay rate.

We finally comment on contributions from loop diagrams.  As discussed 
in Ref.~\cite{Nagata:2013sba}, if the sfermion sector has large flavor 
violation, loop contributions may be significantly enhanced.  For 
instance, large flavor violation in the squark masses can induce 
large corrections to the first and second generation Yukawa couplings, 
as in Fig.~\ref{fig:diagram:yukawa}, and accordingly large corrections 
to the colored Higgs Yukawa couplings.  In some cases, proton decay 
induced through such one-loop diagrams may dominate over the tree-level 
contribution.  The importance of one-loop processes, however, depends 
strongly on the gaugino masses, the structure of sfermion flavor violation, 
GUT-violating threshold corrections from higher dimension operators, 
and so on.  For example, amplitudes of Figs.~\ref{fig:diagram:loop} 
and \ref{fig:diagram:yukawa} are proportional to the gaugino masses, 
so that smaller gaugino masses result in smaller contributions.  Also, 
flavor violation in ${\bf 5}^*$ matter, $\bar{F}(d, l)$, generically 
leads to smaller effects on proton decay than that in ${\bf 10}$ matter, 
$T(q, u, e)$, and, depending on the size of the GUT-violating effects, 
cancellations between contributions from Figs.~\ref{fig:diagram:loop} 
and \ref{fig:diagram:yukawa} may occur.  The general study of all these 
effects is thus highly complicated.  In the explicit analysis in the 
next subsection, we ignore these possible corrections and assume, 
for simplicity, minimal flavor violation for the relevant flavor 
structure.

\subsection{Example:\ mSUGRA-like spectrum}
\label{subsec:mSUGRA}

In this subsection, we present a detailed study of the model described 
here in the case that the supersymmetry breaking masses obey mSUGRA-like 
boundary conditions.  Specifically, we set the following boundary 
conditions for the relevant parameters at the renormalization scale 
$Q_0 = 10^{17}~{\rm GeV}$:
\begin{equation}
  m^2_{T({\bf 10})} = m^2_{\bar{F}({\bf 5}^*)} 
  = \tilde{m}^2\, {\bf 1}_{3\times 3},
\qquad
  m^2_{H_u} = m^2_{H_d} = m^2_{H_C} = m^2_{\bar{H}_C} = \tilde{m}^2,
\label{eq:bc-1}
\end{equation}
\begin{equation}
  M_{\tilde{B}} = M_{\tilde{W}} = M_{\tilde{g}} =  m_{1/2},
\qquad
  \mu = \mu_{H_C},
\qquad
  B = B_{H_C}.
\label{eq:bc-2}
\end{equation}
The $A$ terms are set to zero, and the mass of the uneaten components 
of $\Sigma$ is taken to be $\tilde{m}$.  Here, we have ignored possible 
GUT breaking effects for the above parameters.  While the running mass of 
the wino obtained from these boundary conditions is typically large, we 
assume that its physical mass is $1~{\rm TeV}$ as a result of cancellations 
among various, including threshold, contributions.  (This assumption 
affects renormalization group evolution from TeV to $\tilde{m}$.)  We 
set the Yukawa interactions between the colored Higgs and matter fields 
by matching them with the SSM Yukawa couplings $y_u$ and $y_d$; see 
Ref.~\cite{Nagata:2013sba} for details.  In the analysis below, we treat 
$\tan\beta$ as an input parameter, trading $\mu$ and $B$ with $v$ and 
$\tan\beta$ by the electroweak symmetry breaking condition, as in 
conventional mSUGRA analyses.

\begin{figure}[t!]
\begin{center}
\subfigure[$\tau_p(p \rightarrow \bar{\nu} K^+)$]
  {\includegraphics[clip, width = 0.48 \textwidth]{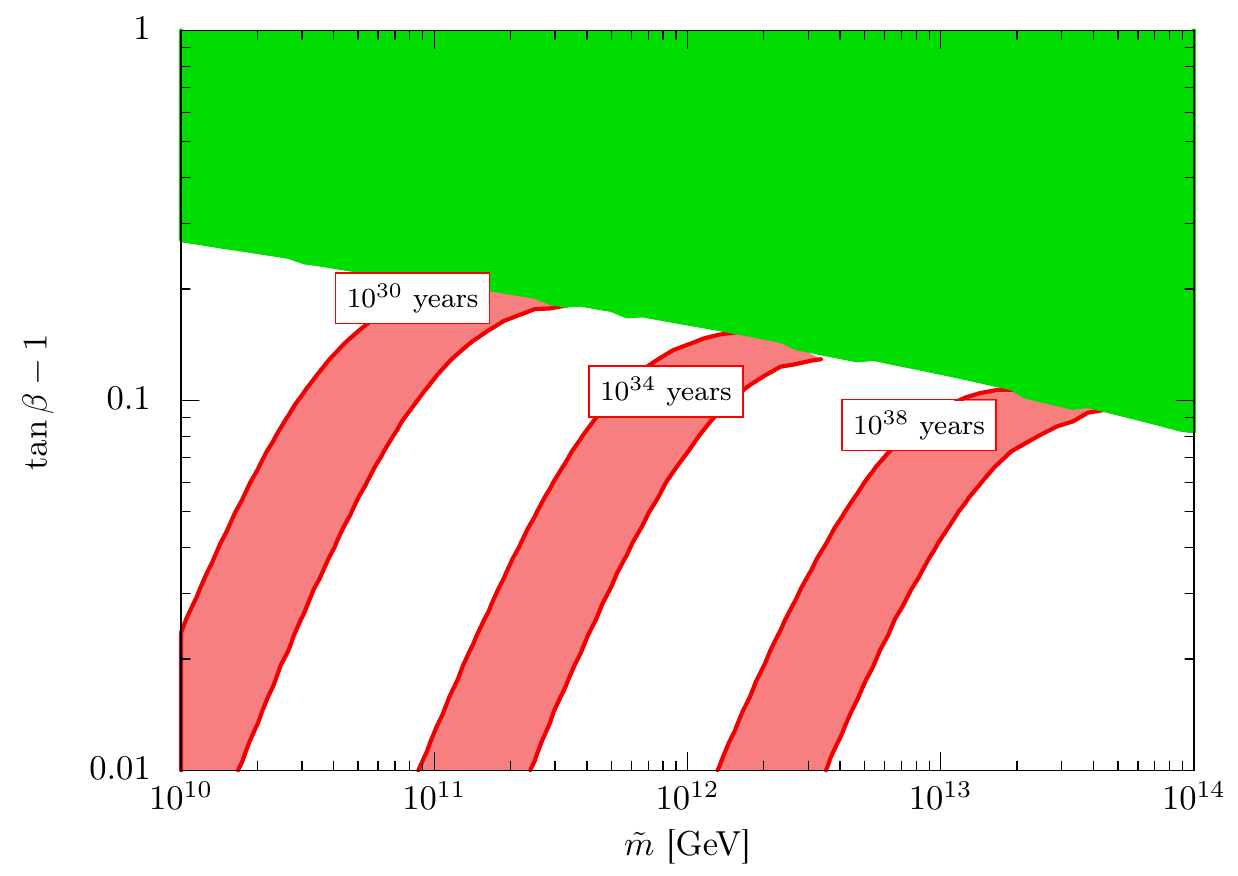}
  \label{fig:mSUGRA_tanb:life}}
\subfigure[Masses in the Higgs sector]
  {\includegraphics[clip, width = 0.48 \textwidth]{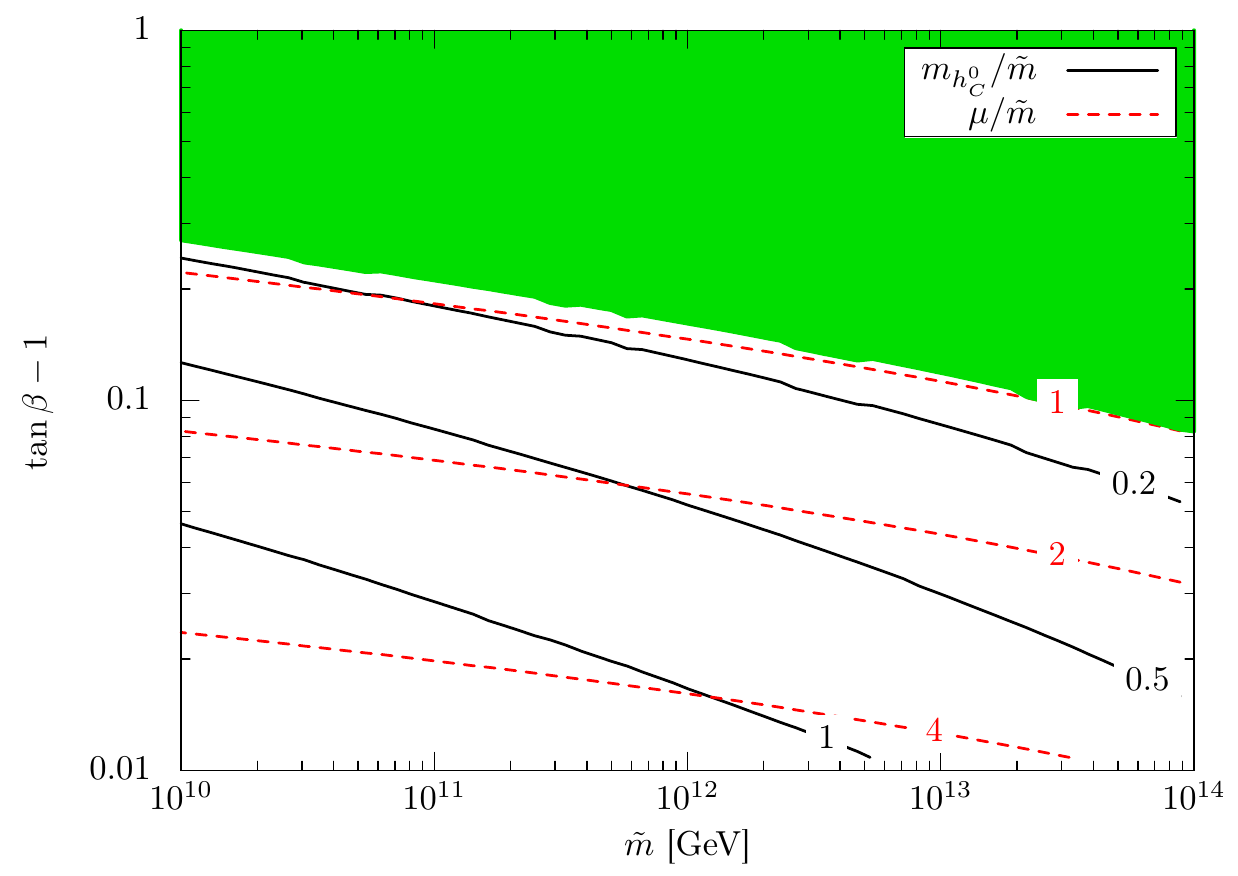}
  \label{fig:mSUGRA_tanb:mass}}
\caption{(a) Proton lifetime for the $\bar{\nu} K^+$ mode and (b) 
 $m_{h^0_C}/\tilde{m}$ (black solid) and $\mu/\tilde{m}$ (red dashed) for 
 the mSUGRA-like boundary conditions in Eqs.~(\ref{eq:bc-1},~\ref{eq:bc-2}) 
 with $m_{1/2} = 0.01 \tilde{m}$.  The boundary conditions in 
 Eq.~(\ref{eq:bc-1}) exclude the green regions at the top.  The 
 red bands for the proton lifetime represent uncertainties from 
 the GUT $CP$ phases discussed in Section~\ref{subsec:p-decay}.}
\label{fig:mSUGRA_tanb}
\end{center}
\end{figure}
In Fig.~\ref{fig:mSUGRA_tanb}, we show the proton lifetime of $p \rightarrow 
\bar{\nu} K^+$ (in \ref{fig:mSUGRA_tanb:life}) and the masses of the lightest 
colored Higgs scalar ${h^0_C}$ and $\mu$ (in \ref{fig:mSUGRA_tanb:mass}) 
in the $\tilde m$-$\tan\beta$ plane.  Here, we have set $m_{1/2}/{\tilde m} 
= 0.01$.  The green shaded region in the upper region of each plot 
represents parameter space in which correct electroweak symmetry breaking 
cannot occur.  We find that $\tan\beta - 1$ cannot be larger than 
$O(0.1)$.  This is because the boundary condition $m_{H_u}^2 = m_{H_d}^2$ 
at $Q = Q_0$ implies $m_{H_u}^2 \approx m_{H_d}^2$ at $Q = \tilde{m}$, 
leading to $\tan\beta \approx 1$; see Eq.~(\ref{eq:tan-beta}). 
(The electroweak symmetry can be broken by $B$.)  Because $\tan\beta 
\sim 1$, essentially all the allowed region with $\tilde{m} \approx 
O(10^9\mbox{--}10^{13}~{\rm GeV})$ and $\mu/\tilde{m} \lesssim 4$ is 
consistent with the observed Higgs boson mass, $m_{h^0} \simeq 125~{\rm GeV}$. 
(For $\mu \gg \tilde{m}$, there could be significant threshold corrections 
to $\lambda$, making it deviate from the condition $\lambda(\tilde{m}) 
\approx 0$.)  The red bands in Fig.~\ref{fig:mSUGRA_tanb:life} 
represent uncertainties from the GUT $CP$ phases discussed in 
Section~\ref{subsec:p-decay}.  In calculating the proton decay rate, 
we have used the CKM matrix in Ref.~\cite{UTfit}, the PDG average 
of the light quark masses~\cite{Beringer:1900zz}, and the four-loop 
renormalization group equations and three-loop decoupling effects from 
heavy quarks~\cite{Chetyrkin:1997sg} to estimate the Yukawa couplings 
of the light quarks.  We have followed Ref.~\cite{Abbott:1980zj} to 
obtain the Wilson operators relevant for the proton decay at the hadronic 
scale, and used matrix elements in Ref.~\cite{Aoki:2013yxa}.

In Fig.~\ref{fig:mSUGRA_tanb:mass}, we see that $\mu/\tilde{m}$ increases 
as we go lower in the plot.  This is because the value of $\tan\beta$ 
is given by
\begin{equation}
  \tan\beta - 1 \approx 
    \frac{m_{H_d}^2-m_{H_u}^2}{2|\mu|^2} \Biggr|_{Q \approx \tilde{m}} 
  \approx O\biggl( \frac{\tilde{m}^2}{|\mu|^2} \biggr),
\label{eq:tanb-1}
\end{equation}
for $|\mu|^2 \gtrsim \tilde{m}^2$.  Note that we need not have an extreme 
fine-tuning between $m_{H_u}^2$ and $m_{H_d}^2$ to obtain $\tan\beta \ll 1$ 
for $|\mu|$ reasonably larger than $\tilde{m}$.  In the figure, we also 
find that the lightest colored Higgs scalar ${h^0_C}$ is a factor of 
a few lighter than the heavier (colored and non-colored) Higgs bosons, 
whose masses are around $|\mu| \sim |\mu_{H_C}|$.  This is because 
$h^0_C$ is almost the GUT partner of the light fine-tuned SM Higgs 
$h^0$, so that its mass is suppressed due to the approximate GUT 
relation between the color-triplet and weak-doublet Higgs mass-squared 
matrices, which is broken here only by the renormalization group running 
effect between $Q_0$ and $\tilde{m}$.  Note that the dominant contribution 
to proton decay comes from exchange of $h^0_C$, with the rate proportional 
to the inverse fourth powers in the mass of $h^0_C$.  This implies that 
the proton decay rate is highly sensitive to possible GUT-violating 
threshold corrections at $Q_0$.  For example, an $O(10\%)$ violating 
of, e.g., the relation $\mu = \mu_{H_C}$ or $m^2_{H_u,H_d} = 
m^2_{H_C,\bar{H}_C}$ could lead to a change of the proton decay 
rate by a couple of orders of magnitude.

\begin{figure}[t!]
\begin{center}
\subfigure[$\tau_p(p \rightarrow \bar{\nu} K^+)$]
  {\includegraphics[clip, width = 0.48 \textwidth]{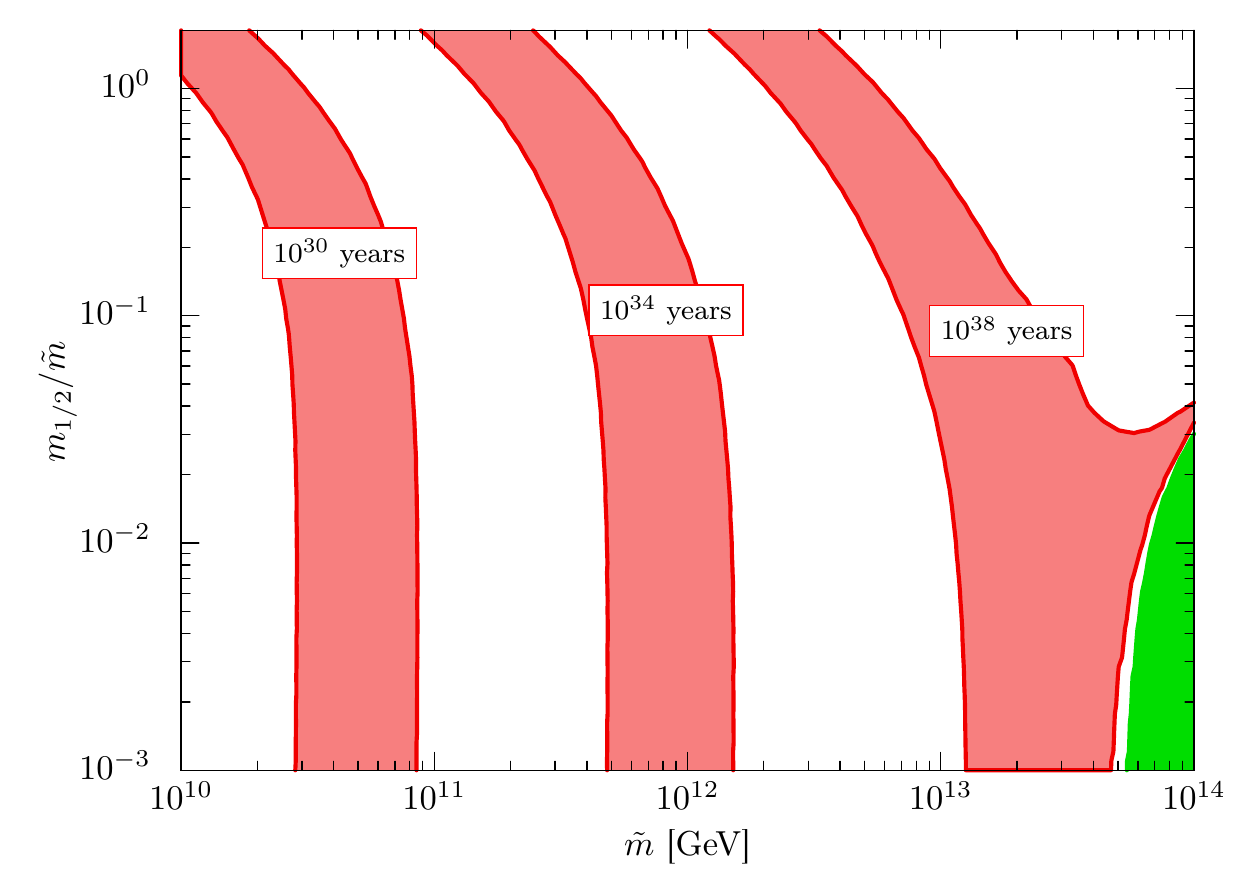}
  \label{fig:mSUGRA_R:life}}
\subfigure[Masses in the Higgs sector]
  {\includegraphics[clip, width = 0.48 \textwidth]{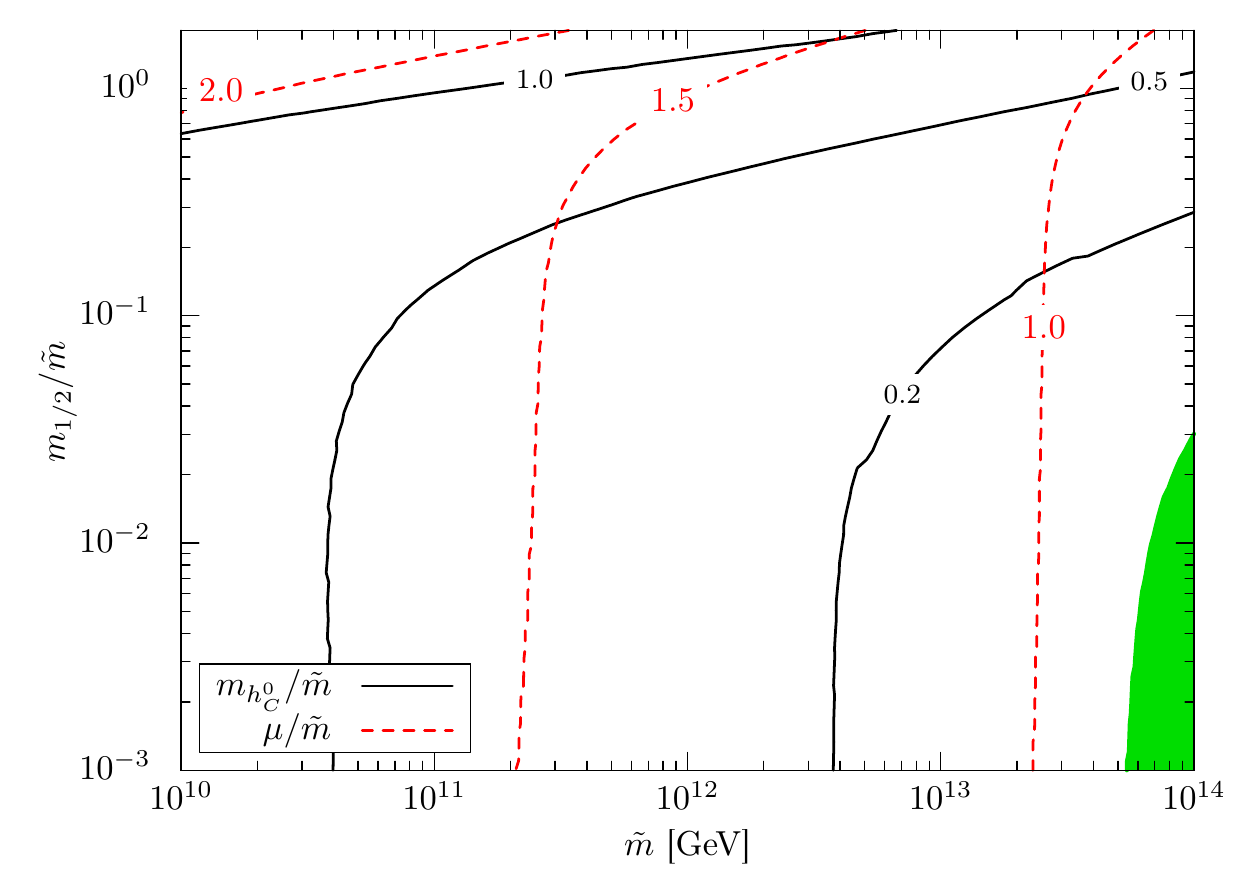}
  \label{fig:mSUGRA_R:mass}}
\caption{Same quantities as in Fig. \ref{fig:mSUGRA_tanb} plotted in 
 the $\tilde{m}$-$m_{1/2}/\tilde{m}$ plane. We have set $\tan\beta = 1.1$.}
\label{fig:mSUGRA_R}
\end{center}
\end{figure}
In Fig.~\ref{fig:mSUGRA_R}, we show the same quantities as in 
Fig.~\ref{fig:mSUGRA_tanb} in the $\tilde{m}$-$m_{1/2}/\tilde{m}$ 
plane by taking $\tan\beta = 1.1$.  As we increase $m_{1/2}/\tilde{m}$, 
the mass of $h^0_C$ becomes larger and, accordingly, the lifetime of 
the proton for fixed $\tilde{m}$ increases.  This is because larger 
$m_{1/2}/\tilde{m}$ leads to a larger violation of the GUT relation 
between the mass-squared matrices for the color-triplet and weak-doublet 
Higgses, so that electroweak fine-tuning for the mass of $h^0$ yields 
less suppression for the mass of $h^0_C$.

\section{ISS as the Origin of Scales for New Physics}
\label{sec:scales}

In this section, we argue that the mass scale $\tilde{m}$ in ISS may 
provide the origin of a variety of new physics occurring at intermediate 
scales, Eq.~(\ref{eq:m-tilde}).  Specifically, we consider the heavy 
mass scale for seesaw neutrino masses, the axion decay constant, and 
the inflaton mass as originating from $\tilde{m}$.  The discussions 
here are not contingent on the specific model presented in the previous 
section or in Ref.~\cite{Hall:2013eko}, and apply more generally to 
a large classes of ISS models.  Also, all the mechanisms described below 
need not be realized simultaneously; one or more of the mass scales 
appearing in these phenomena may originate from other physics.

\subsection{Seesaw neutrino masses}
\label{subsec:neutrino}

The simplest understanding of small neutrino masses follows from having 
lepton number broken at a very high scale, $M_L$.  At energies below 
$M_L$, lepton number becomes an accidental symmetry of interactions 
up to dimension~4, yielding Majorana neutrino masses at dimension~5 
via $ll\, hh/M_L$.  Within ISS it is natural to associate $M_L$ with 
$\tilde{m}$, since this is the only mass scale below the cutoff, 
giving neutrino masses of $m_\nu \sim v^2/\tilde{m}$.

We can implement this by introducing $SU(5)$ singlet right-handed 
neutrino superfields, $N$, neutral under $U(1)_R$, so that the K\"{a}hler 
potential contains $c_\nu NN/2$, with $c_\nu$ being $O(1)$ coefficients. 
Once supersymmetry is broken, the supergravity interactions generate 
an effective superpotential $W_{\rm eff}^\nu = \tilde{m}_\nu NN/2$, 
where $\tilde{m}_\nu$ is a $3 \times 3$ matrix in flavor space with 
entries order $\tilde{m}$.  Introducing a $3 \times 3$ Yukawa coupling 
matrix $y_\nu$ via the superpotential interaction $y_\nu N \bar{F} H$ 
leads to a light neutrino mass matrix
\begin{equation}
  m_\nu = \left( y_\nu^T\, \frac{1}{\tilde{m}_\nu}\, y_\nu \right) v^2.
\label{eq:mnu}
\end{equation}
For example, with $\tilde{m} = 10^{13}~{\rm GeV}$, the observed neutrino 
masses result from $y_\nu$ entries of order $0.1$.

Previously we have used a $U(1)_R$ symmetry with charges $R=0$ for 
$H, \bar{H}, \Sigma$ and $R=1$ for $T, \bar{F}$.  This symmetry does 
not work in the present case, since the $y_\nu N \bar{F} H$ couplings 
would then imply $N$'s having $R=1$, so that $K \supset c_\nu NN/2$ 
cannot be written while the $N^2$ terms are allowed in the superpotential. 
Assuming that the supersymmetry breaking field $X$ is neutral under 
it, we find a unique flavor-independent $R$ symmetry that allows both 
$c_\nu$ and $y_\nu$ to be non-zero:\ $R' = R - X/5$, where $U(1)_X$ 
is the Abelian generator that appears in $SO(10)/SU(5)$ and has charges 
$T(1), \bar{F}(-3), N(5), H(-2), \bar{H}(2), \Sigma(0)$, and we have 
chosen $R=1$ for $N$.  Imposing this $U(1)_{R'}$ symmetry yields a 
theory with the general structure
\begin{align}
  \tilde{K} &= \Lambda^2\, f\Bigl( \frac{\Sigma}{\Lambda}, 
    \frac{H\bar{H}}{\Lambda^2}, \frac{NN}{\Lambda^2} \Bigr),
\label{eq:model-Knu}\\
  W &= y_U\Bigl( \frac{\Sigma}{\Lambda} \Bigr)\, T T H 
    + y_D\Bigl( \frac{\Sigma}{\Lambda} \Bigr)\, T \bar{F} \bar{H}
    + y_\nu\Bigl( \frac{\Sigma}{\Lambda} \Bigr)\, N \bar{F} H,
\label{eq:model-Wnu}
\end{align}
leading to the neutrino mass matrix in Eq.~(\ref{eq:mnu}).  Here, $\tilde{K}$ 
is the holomorphic part of the K\"{a}hler potential, and we have imposed 
matter parity under which $\Sigma, H, \bar{H}$ are even while $T, \bar{F}, 
N$ are odd.%
\footnote{An introduction of separate matter parity can be avoided 
 if we consider an $R$ symmetry under which the supersymmetry breaking 
 field $X$ is charged.  In this case, the right-handed neutrino masses 
 may be generated by operators like $K \supset X^\dagger NN/2$, and 
 $R$-parity may arise as an unbroken ${\bf Z}_2$ subgroup of the $R$ 
 symmetry.}

It is interesting to note that values of the right-handed neutrino 
masses implied by the above mechanism are consistent with thermal 
leptogenesis, which works for a wide range of conditions after 
inflation if the lightest right-handed neutrino is heavier 
than about $10^9~{\rm GeV}$ for hierarchical right-handed 
neutrinos~\cite{Giudice:2003jh}.

\subsection{Axion}
\label{subsec:axion}

One of the major problems in the SM is the strong $CP$ problem.  A 
promising solution is to introduce an anomalous Peccei-Quinn (PQ) 
symmetry spontaneously broken at a scale $f_a$, leading to an axion 
field with decay constant $f_a$.  Here we consider that the scale of 
$f_a$ is given by $\tilde{m}$, and present a simple model realizing 
this idea.  For a different implementation of a similar setup, in 
which $f_a$ is related to $\tilde{m}$, see Ref.~\cite{Barger:2004sf}.

We consider the superpotential of the form
\begin{equation}
  W \supset c\, S Q \bar{Q} + c'\, S^2 \bar{S}.
\label{eq:W_PQ}
\end{equation}
Here, $c$ and $c'$ are coefficients of order unity, and chiral 
superfields $S$ (which will be identified as the PQ-breaking field), 
$\bar{S}, Q$, and $\bar{Q}$ have the $U(1)$ PQ charges
\begin{equation}
  S(1),\quad \bar{S}(-2),\quad Q\bar{Q}(-1).
\label{eq:PQ}
\end{equation}
The superpotential of the above form may be used to build a variety of 
axion models, including DFSZ-type models in which a part of $Q$ and 
$\bar{Q}$ may be identified with the SSM Higgs doublets.  Here we choose 
the following simple $SU(5)$ representation
\begin{equation}
  S({\bf 1}),\quad \bar{S}({\bf 1}),\quad Q({\bf 5}),\quad \bar{Q}({\bf 5}^*).
\label{eq:ax-SU5}
\end{equation}
Since $Q$ and $\bar{Q}$ comprise complete $SU(5)$ multiplets, gauge coupling 
unification is preserved.  This simple choice also guarantees that the 
so-called domain wall number $N_{\rm DW}$ is unity, which allows us to 
avoid stringent cosmological constraints as discussed below.

Once supersymmetry is broken, the $S$ field may have a negative soft 
supersymmetry-breaking mass-squared of order $\tilde{m}^2$:
\begin{equation}
  V \supset - m_S^2 S^{\dagger} S,
\qquad
  m_S^2 \sim O(\tilde{m}^2).
\label{eq:mS2}
\end{equation}
Indeed, this negative mass-squared may be induced radiatively through 
renormalization group running from $M_*$ ($\sim \Lambda \sim M_{\rm Pl}$) 
to the scale $\tilde{m}$, starting from the boundary condition 
that all the fields have positive soft squared masses at $M_*$, 
in which case the soft supersymmetry-breaking squared masses for 
$\bar{S}$, $Q$, and $\bar{Q}$ will be positive.  The potential given 
by Eqs.~(\ref{eq:W_PQ},~\ref{eq:mS2}) leads to a vacuum at
\begin{equation}
  \langle S \rangle = f_a \sim O(\tilde{m}),
\qquad
  \langle \bar{S} \rangle = \langle Q \rangle = \langle \bar{Q} \rangle = 0,
\label{eq:PQ-vacuum}
\end{equation}
breaking the PQ symmetry at $\sim \tilde{m}$.  As a result, all the 
particles in the $S, \bar{S}, Q, \bar{Q}$ multiplets receive masses of 
order $\tilde{m}$ except for the light Nambu-Goldstone axion field arising 
from $S$, whose decay constant is given by $\langle S \rangle = f_a$.

The recent discovery of the $B$-mode polarization in cosmic microwave 
background radiation by BICEP2 collaboration~\cite{Ade:2014xna}, 
$r = 0.2^{+0.07}_{-0.05}$, suggests that the scale of inflation is 
large:
\begin{equation}
  H_I \simeq 7.8 \times 10^{13} ~{\rm GeV} 
    \left( \frac{r}{0.1} \right)^{1/2},
\label{eq:H_inf}
\end{equation}
where $H_I$ is the Hubble parameter during inflation.  This has significant 
impacts on axion models.  If the PQ symmetry is broken before inflation, 
the light axion field produces isocurvature perturbation.  With the 
large inflation scale in Eq.~(\ref{eq:H_inf}), this case is excluded 
by observation~\cite{Fox:2004kb}, unless the dynamics associated 
with the PQ symmetry breaking is somewhat complicated, e.g., if $f_a$ 
is much larger~\cite{Linde:1991km} or if the axion is heavier than 
$H_I$~\cite{Jeong:2013xta} during inflation due to nontrivial dynamics.

We thus consider here the case in which the PQ symmetry is broken after 
the end of inflation.  In this case, topological objects formed associated 
with PQ symmetry breaking, in particular domain walls, may give serious 
cosmological problems~\cite{Sikivie:1982qv}.  In the model presented 
above, however, the domain wall number is unity, $N_{\rm DW} = 1$, 
so that domain walls, which are disk-like objects bounded by strings, 
are unstable~\cite{Vilenkin:1982ks}.  The decay of these domain walls 
produces axion particles, but a detailed lattice simulation indicates 
that the value of $f_a \lesssim \mbox{a few} \times 10^{10}~{\rm 
GeV}$ is consistent with the current observation~\cite{Hiramatsu:2012gg}. 
(A slightly weaker estimate of $f_a \lesssim 10^{11}~{\rm GeV}$, coming 
from the misalignment mechanism of dark matter production, is implied 
by the analysis in Ref.~\cite{Chang:1998tb}.)  Together with the lower 
bound on $f_a$ from stellar physics (for reviews on axion physics, see 
e.g.~\cite{Raffelt:1999tx}), we find that
\begin{equation}
  f_a \approx O(10^9\mbox{--}10^{11}~{\rm GeV}),
\label{eq:f_a}
\end{equation}
gives consistent phenomenology.  (We expect that the axino and saxion do 
not cause cosmological problems in the present model, since they are heavy 
with masses of order $\tilde{m}$.  The $Q$ and $\bar{Q}$ states may also 
be made to decay by coupling them with SM particles, without violating 
the PQ symmetry.)  The value of $f_a$ in Eq.~(\ref{eq:f_a}) can be easily 
obtained with $\tilde{m}$ in the ISS range, suggested by the observed 
Higgs boson mass.

\subsection{Inflation}
\label{subsec:inflation}

A very simple inflation model is given by the following potential for the 
inflaton $\phi$:
\begin{equation}
  V(\phi) = \frac{m_\phi^2}{2}\, \phi^2.
\label{eq:V_inf}
\end{equation}
Interestingly, this simple potential gives a good agreement with the 
observations of the scalar spectral index $n_s$ by Planck~\cite{Ade:2013zuv} 
and the tensor-to-scalar ration $r$ by BICEP2~\cite{Ade:2014xna} for
\begin{equation}
  m_\phi \simeq 10^{13}~{\rm GeV}.
\label{eq:m_inf}
\end{equation}
It is, therefore, interesting to identify $m_\phi$ as $\tilde{m}$, 
which is roughly in the same energy range.%
\footnote{While completing this paper a similar observation was made 
 in Ref.~\cite{Ibanez:2014zsa}.}

The construction of a complete inflation model in supergravity realizing 
the above idea, however, is challenging, since the value of field $\phi$ 
during inflation is beyond the reduced Planck scale $M_{\rm Pl}$, where 
the scalar potential obtained from supergravity loses theoretical control. 
Moreover, depending on the mechanism of how the supersymmetry breaking 
masses are generated, the soft supersymmetry breaking mass for $\phi$ may be 
shut off above some scale, e.g., $M_*$.  One possibility is to use a shift 
symmetry on $\phi$ along the lines of, e.g., Ref.~\cite{Kawasaki:2000yn}, 
but the construction of an explicit model seems nontrivial.  Another 
possibility is that the apparent obstruction in supergravity of having 
flat potential beyond $\phi \approx M_{\rm Pl}$ is not warranted, as has 
been discussed, e.g., in Ref.~\cite{Kaloper:2008fb} and more recently 
in Ref.~\cite{Kehagias:2014wza}.

An alternative direction for realizing the idea of connecting the ISS scale 
with inflation is to use the constant term in the superpotential, $W_0$, 
needed to cancel the cosmological constant.  If we assume that the 
superpartner mass scale $\tilde{m}$ is generated by some mediation 
mechanism at $M_*$, the $F$-term VEV for the supersymmetry breaking field 
is given by $F \sim \tilde{m} M_*$.  This implies that $W_0 \sim \tilde{m} 
M_* M_{\rm Pl}$.  Taking $M_* \sim M_{\rm Pl}$, this scale is thus
\begin{equation}
  W_0^{1/3} \sim 10^{16} 
    \left( \frac{\tilde{m}}{10^{12}~{\rm GeV}} \right)^{1/3} {\rm GeV},
\label{eq:W_0}
\end{equation}
which is very close to the energy scale during inflation $V_I^{1/4} \simeq 
2 \times 10^{16}~{\rm GeV}$ suggested by the BICEP2 data.  Inflation, 
therefore, may occur associated with the dynamics generating this constant 
term, for example through some gaugino condensations, along the lines 
of, e.g., Ref.~\cite{Adams:1992bn}.  We leave explorations of explicit 
inflation models in the ISS framework to future work.

\section{Summary}
\label{sec:summary}

We have explored supersymmetric grand unified theories that have a single 
scale, that of supersymmetry breaking, determined by the value of the 
Higgs boson mass to be in the intermediate range of  $\tilde{m} \sim 
10^9\mbox{--}10^{13}~{\rm GeV}$.  Mass terms for the $SU(5)$ Higgs 
multiplets, $\Sigma, H, \bar{H}$ are generated at $\tilde{m}$ in the 
same way that in minimal supersymmetric models the Higgs mass parameter 
$\mu$ can arise at the supersymmetry breaking scale.  However, unlike 
electroweak breaking in these minimal models, the breaking of the 
unified symmetry by $\Sigma$ occurs at a scale parametrically higher 
than $\tilde{m}$, close to the cutoff scale of the theory.

\begin{figure}[t!]
\begin{center}
  {\includegraphics[clip]{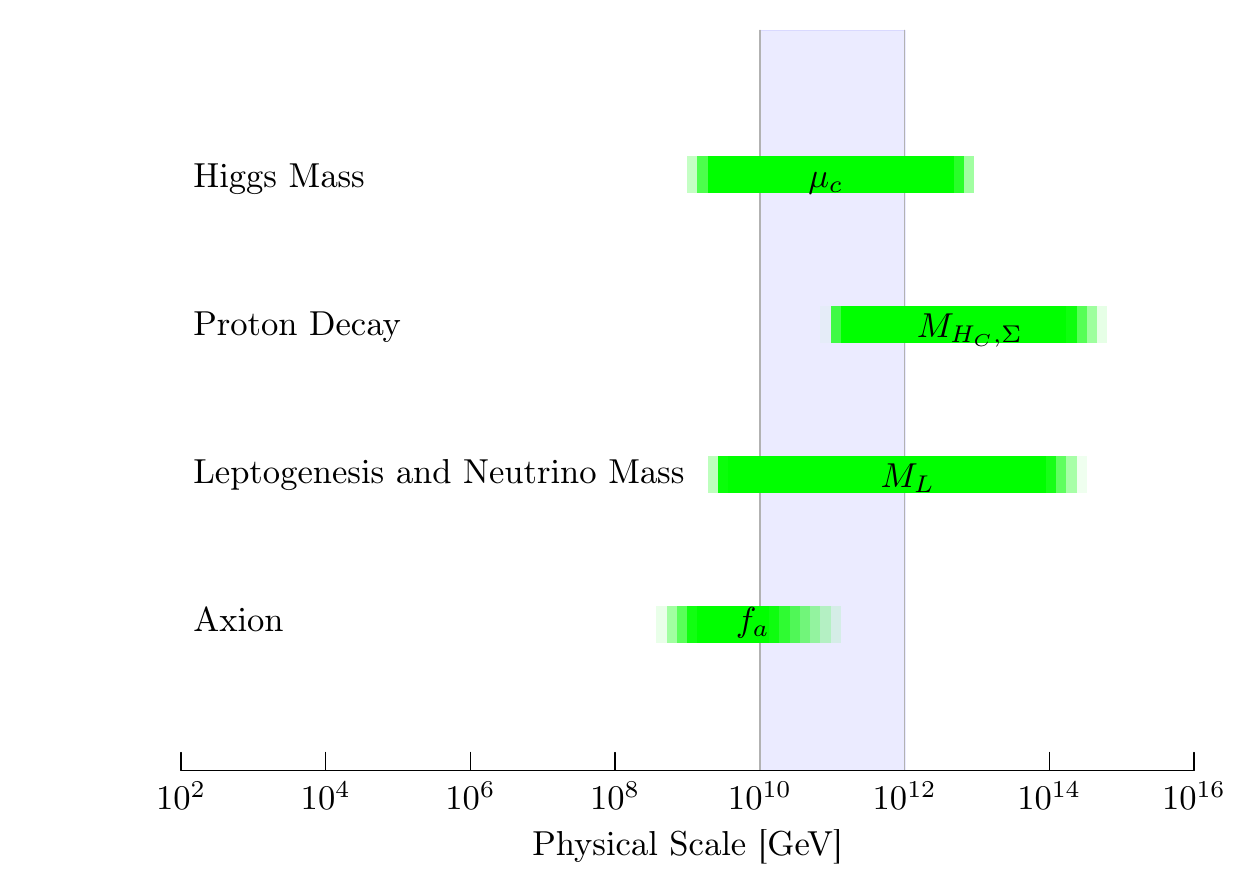}}
\caption{The experimentally allowed ranges of four key mass scales:\ 
 $\mu_c$ (the scale at which the SM Higgs quartic coupling vanishes); 
 $M_{H_C,\Sigma}$ (the masses of the $H_c$ and $\Sigma$ states in the 
 ISS model of Section~\ref{sec:model}); $M_L$ (the scale of lepton number 
 violation for seesaw neutrino masses and leptogenesis); and $f_a$ (the 
 axion decay constant in minimal models that solve the strong $CP$ problem). 
 All are consistent with ISS, with supersymmetry breaking centered around 
 the shaded region.}
\label{fig:summ}
\end{center}
\end{figure}

A variety of diverse physics can be described by such GUTs with ISS, 
as illustrated in Fig.~\ref{fig:summ}.  For a wide range of parameters, 
the SM Higgs quartic coupling is constrained to be small at $\tilde{m}$; 
indeed we determine the allowed range of $\tilde{m}$ by using the Higgs 
mass as input as shown in Fig.~\ref{fig:scale}.  The result is illustrated 
by the upper horizontal green bar in Fig.~\ref{fig:summ}, showing the 
range of the scale $\mu_c$ where the quartic coupling vanishes in the 
SM (possibly augmented by a TeV wino for dark matter).

In the minimal ISS model, introduced and studied in depth in 
Section~\ref{sec:model}, proton decay is induced by both the tree-level 
exchange of the colored triplet $SU(5)$ partner of the Higgs boson 
$H_C$, of mass $M_{H_C}$, and by the exchange of the GUT gauge bosons 
$X$.   The mass of $H_C$ is expected to be comparable to the mass 
of the uneaten states in $\Sigma$, $M_\Sigma$, and the experimental 
constraint on these masses is shown in the second horizontal green bar 
of Fig.~\ref{fig:summ}.  The lower end of the range results from the 
limit on $p \rightarrow \bar{\nu} K^+$ from $H_C$ exchange, while the 
upper end of the range arises from the limit on $p \rightarrow e^+ \pi^0$ 
from $X$ exchange; the mass of $X$ being sensitive to $M_\Sigma$ via 
gauge coupling unification.  Even though there are order unity couplings 
that lead to differences between $\mu_c$ and $M_{H_C, \Sigma}$, it 
is important for the consistency of the theory that the ranges of the 
top two green bars overlap.  While the presence of $\Sigma$ states at 
$\tilde{m}$ solves the proton decay problem of non-supersymmetric $SU(5)$, 
having $H_C$ states at $\tilde{m}$ does not introduce a new proton decay 
problem, but offers the possibility of a signal.  The precision of gauge 
coupling unification is further enhanced if dark matter is environmentally 
selected by fine-tuning the wino mass to the TeV region.

The basic model of Section~\ref{sec:model} leaves open two key questions, 
the origin of neutrino masses and inflation.  Seesaw neutrino masses 
occur very naturally in our framework as the lepton number violating 
mass for the right-handed neutrinos, $M_L$, can arise from the same 
mechanism that generates the masses for $\Sigma, H, \bar{H}$.  The 
experimentally allowed range for $M_L$ is shown by the third horizontal 
bar in Fig.~\ref{fig:summ}.  The upper end of the range arises from 
the need to explain the size of the atmospheric neutrino oscillation, 
and is shown for neutrino Yukawa couplings of order unity, while the 
lower end arises from the requirement of a leptogenesis origin for 
the cosmological baryon asymmetry.  Note that leptogenesis also requires 
that $M_L$ be less than the reheat temperature after inflation, so that 
the upper bound on $M_L$ may be lower than shown.

Recent data from BICEP2 indicates that the scale of the vacuum energy 
that drives inflation is $\simeq 2 \times 10^{16}~{\rm GeV}$.  However, 
this need not be a Lagrangian mass scale; for example, for an inflation 
potential $m_\phi^2 \phi^2/2$ the required inflaton mass is $m_\phi 
\simeq 10^{13}~{\rm GeV}$.  We do not show this in Fig.~\ref{fig:summ} 
because it is specific to this particular potential.    However, it is 
certainly consistent with the masses $\mu_c, M_{H_c,\Sigma}, M_L$, so 
we may expect that this also arises from $\tilde{m}$.

Finally, the axion is the most promising solution to the strong $CP$ 
problem, and may also account for dark matter.  The large value of the 
Hubble parameter during inflation indicated by the BICEP2 data, implies 
that the simplest axion models having PQ symmetry broken during inflation 
are excluded.  In Fig.~\ref{fig:summ} we therefore show the experimentally 
allowed range of the axion decay constant in theories having a PQ phase 
transition after inflation.  The upper limit arises from overclosure 
by axions, and the lower limit from axion emission from supernovae and 
white dwarfs.  Again, from Fig.~\ref{fig:summ} we notice a remarkable 
consistency between the mass scales required for very different 
physics; in ISS these masses are not precisely equal, but may all 
arise from $\tilde{m}$, the scale of supersymmetry breaking.

\section*{Acknowledgments}

This work was supported in part by the Director, Office of Science, Office 
of High Energy and Nuclear Physics, of the US Department of Energy under 
Contract DE-AC02-05CH11231 and in part by the National Science Foundation 
under grants PHY-0855653 and PHY-1214644.

\end{document}